\documentclass[%
 aip,
 amsmath,amssymb,
 reprint,%
]{revtex4-1}

\usepackage{graphicx}
\usepackage{dcolumn}
\usepackage{bm}

\usepackage[utf8]{inputenc}
\usepackage[T1]{fontenc}
\usepackage{mathptmx}
\usepackage{etoolbox}

\usepackage{comment}

\usepackage{tikz}
\usetikzlibrary{positioning,fit,calc}
\usetikzlibrary{decorations.markings}
\usetikzlibrary{decorations.pathreplacing,calligraphy}
\usetikzlibrary{plotmarks}

\makeatletter
\def\@email#1#2{%
 \endgroup
 \patchcmd{\titleblock@produce}
  {\frontmatter@RRAPformat}
  {\frontmatter@RRAPformat{\produce@RRAP{*#1\href{mailto:#2}{#2}}}\frontmatter@RRAPformat}
  {}{}
}%
\makeatother

\begin{document}

\preprint{AIP/123-QED}

\title{Strong collisionless coupling between an unmagnetized driver plasma and a magnetized background plasma}

\author{F. D. Cruz}
 \email{filipe.d.cruz@tecnico.ulisboa.pt}
 \affiliation{GoLP/Instituto de Plasmas e Fus\~{a}o Nuclear, Instituto Superior T\'{e}cnico, Universidade de Lisboa, 1049-001 Lisboa, Portugal}

\author{D. B. Schaeffer}
 \affiliation{Department of Physics and Astronomy, University of California --- Los Angeles, Los Angeles, California 90095, USA}

\author{F. Cruz}
\affiliation{GoLP/Instituto de Plasmas e Fus\~{a}o Nuclear, Instituto Superior T\'{e}cnico, Universidade de Lisboa, 1049-001 Lisboa, Portugal}
\affiliation{Inductiva Research Labs, Rua da Prata 80, 1100-420 Lisboa, Portugal}

\author{L. O. Silva}
\affiliation{GoLP/Instituto de Plasmas e Fus\~{a}o Nuclear, Instituto Superior T\'{e}cnico, Universidade de Lisboa, 1049-001 Lisboa, Portugal}

\date{\today}

\begin{abstract}
Fast-exploding plasmas traveling though magnetized, collisionless plasmas can occur in a variety of physical systems, such as supernova remnants, coronal mass ejections, and laser-driven laboratory experiments. To study these systems, it is important to understand the coupling process between the plasmas. In this work, we develop a semi-analytical model of the parameters that characterize the strong collisionless coupling between an unmagnetized driver plasma and a uniformly and perpendicularly magnetized background plasma. In particular, we derive analytical expressions that describe the characteristic diamagnetic cavity and magnetic compression of these systems, such as their corresponding velocities, the compression ratio, and the maximum size of the cavity. The semi-analytical model is compared with collisionless 1D particle-in-cell simulations and experimental results with laser-driven plasmas, showing good agreement. The model allows us to provide bounds for parameters that are otherwise difficult to diagnose in experiments with similar setups.
\end{abstract}

\maketitle

\section{\label{sec:introduction} Introduction}

\par Interactions between fast-expanding driver plasmas with magnetized background plasmas are commonly observed in astrophysical and space phenomena. Such examples include the interaction of stellar material with Earth's magnetosphere~\cite{Burlaga2001} and with the surrounding medium in supernova remnants~\cite{Spicer1990}, the formation of cometary plasma tails due to the solar wind~\cite{Mendis1977}, and artificial explosions in the Earth's upper atmosphere~\cite{Krimigis1982,Johnson1992,Palmer2006}. To model the physics behind these phenomena, it is important to comprehend the interaction processes between the driver and background plasmas.%

\par This interaction can be described through the coupling, \textit{i.e.}, the energy and momentum transfer efficiency between the driver and the background plasmas~\cite{Hewett2011}. In the typical rarefied environments of astrophysical and space systems, collisions between particles are ineffective. For these collisionless processes, it is the electromagnetic fields that determine the physics of the coupling between the two plasmas~\cite{Winske2007}.%

\par A common feature of these systems is the formation of a diamagnetic cavity~\cite{Gurnett1986,Ip1987,Bernhardt1992,Zakharov1999,Collette2010,Goetz2016,Schaeffer2018,Winske2019}. In the interaction region between the driver and the background, the electrons are magnetized while the ions are effectively unmagnetized. The resultant $\mathbf{E}\times\mathbf{B}$ electron drift leads to currents that expel the magnetic field within the driver region while compressing it at the driver's edge~\cite{Ripin1993,VanZeeland2004,Hewett2011}.%

\par Many previous analytical and numerical studies on the coupling between a driver and a magnetized plasma have focused on estimating the maximum size of the diamagnetic cavity~\cite{VanZeeland2004,Bonde2018I,Bonde2018II,Ari2021,Behera2021}. From energy conservation arguments, this size was calculated for sub-Alfvénic (Alfvénic Mach number $M_A < 1$)~\cite{Ripin1993} and super-Alfvénic ($M_A > 1$) regimes~\cite{Winske2007,Clark2014}. By assuming hybrid models, some studies described the electric field of the system~\cite{Bondarenko2017}, while others determined the conditions where the plasmas fail to couple with each other~\cite{Hewett2011}. Some attempts were also made to estimate the level of magnetic compression that results from the coupling~\cite{Clark1973,Cairns1994,Zakharov1999,Clark2013,Everson2016}.%

\par Despite the substantial efforts in studying the coupling for these systems, it has been difficult to experimentally verify the obtained models with \textit{in situ} observations from spacecraft, due to their limited control, data, and reproducibility. Motivated by these challenges, in recent decades, multiple scaled laboratory experiments have explored the interaction of laser-produced driver plasmas with magnetized background plasmas in collisionless regimes~\cite{Drake2000,Niemann2014,Bondarenko2017}. Some of these experiments focused on validating analytical and numerical models by measuring the size of the diamagnetic cavity for multiple parameters~\cite{Zakharov1999,VanZeeland2004,Bonde2018II}, and by exploring the electromagnetic fields of the system~\cite{Bondarenko2017,Schaeffer2018}. Other experiments, however, have focused on improving the momentum and energy transfer from the driver to the background and identified different regimes of coupling strength, from weak to strong coupling~\cite{Niemann2013,Schaeffer2015}.%

\par The coupling study presented in this work is motivated by recent experiments to study laboratory ion-scale magnetospheres \cite{Schaeffer2022} performed on the Large Plasma Device (LAPD) at the University of California---Los Angeles~\cite{LAPD}. In these experiments, fast collisionless plasma flows generated by high-repetition-rate lasers were driven against a magnetized background plasma and a dipolar magnetic field. Under this configuration, the main characteristics of ion-scale magnetospheres were observed, and additional particle-in-cell (PIC) simulations supported the analysis of the experimental results~\cite{Cruz2022}. A key component of these experiments is the initial interaction between the laser-driven plasma and background plasma; however, this interaction is difficult to diagnose directly and was limited to measurements of the magnetic field.  A better model of the coupling mechanics is needed to both constrain plasma parameters that cannot be directly measured, and to predict the parameters necessary to achieve different magnetosphere regimes.%

\par In this work, we obtain analytical expressions for multiple parameters that describe the strong coupling between a uniform unmagnetized driver plasma and a uniform perpendicular magnetized background plasma, in non-relativistic conditions. In particular, we derive expressions for the velocities of the upstream magnetic cavity and the downstream magnetic compression, the compression ratio, and the maximum size of the cavity. The expressions for these coupling parameters are consistent with 1D PIC simulations with long plasmas and low electron and ion temperatures. These coupling parameters can be directly obtained from standard magnetic field diagnostics, allowing us to evaluate the coupling between the plasmas, and can be used as a benchmark for the initial conditions of the system. The derived expressions could then be used to design future experiments. We also check the validity of the coupling study against the experimental data of ion-scale magnetospheres and other experiments.%

\par This paper is organized as follows. In Sec.~\ref{sec:review}, we review the previous results with experimental and numerical ion-scale magnetospheres~\cite{Schaeffer2022,Cruz2022} and detail the main motivation for the coupling study. In Sec.~\ref{sec:simulations}, we outline the configuration and parameters used to describe the system with a uniform driver flowing against a uniform magnetized background plasma. Using PIC simulations, we provide an overview of the evolution of these systems and define the parameters that best describe the coupling between the plasmas. In Sec.~\ref{sec:derivation}, we derive analytical expressions for these coupling parameters by using relationships from jump conditions and conservation arguments. In Sec.~\ref{sec:comparison}, we compare these expressions with PIC simulations for scans of the magnetic field, driver density, and ion mass. We also compare the derived size of the magnetic cavity with the simulations. In Sec.~\ref{sec:applications}, we apply the coupling study to experimental data of ion-scale magnetospheres. Finally, in Sec.~\ref{sec:conclusions}, we outline the conclusions of this work.%

\section{\label{sec:review} Motivation}

\par The motivation for this study was a series of experiments on laboratory ion-scale magnetospheres~\cite{Schaeffer2022}. The experiments aimed to demonstrate ion-scale magnetosphere formation using a laser-driven plasma expanding into a dipolar magnetic field, embedded in a magnetized background plasma. A key component of these experiments was energy and momentum coupling between the laser-produced plasma and the background plasma, both of which were found to play an important role in the magnetosphere formation.

\par
\begin{figure}[!h]
    \centering
    \includegraphics[width=0.9\columnwidth]{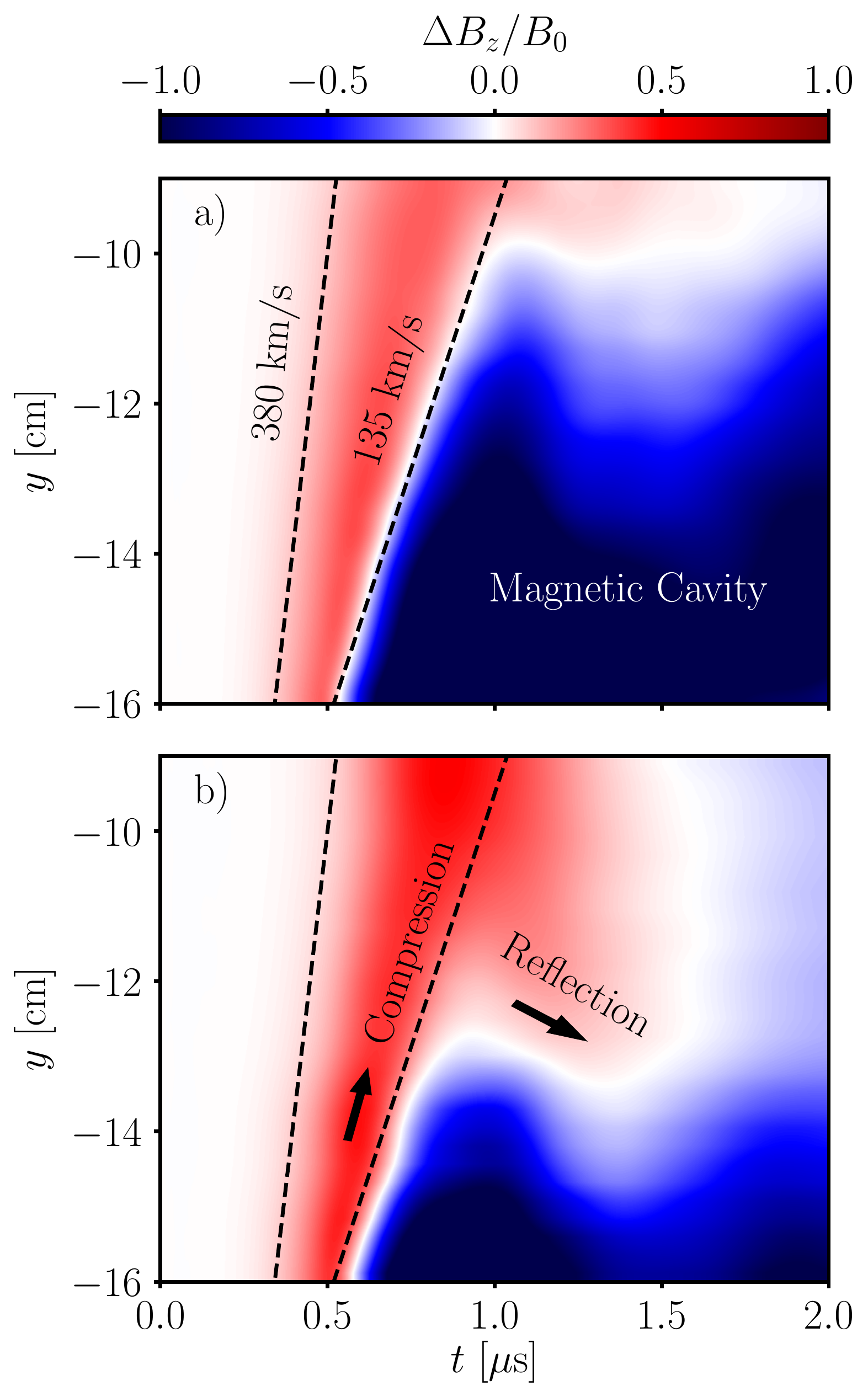}
    \vspace{-2ex}
    \caption{LAPD experimental results with laboratory ion-scale magnetospheres for the evolution of the variation of the magnetic field $\Delta B_z$, along the symmetry axis $x=z=0$, in the ``dayside'' region of the magnetosphere~\cite{Schaeffer2022}. The dipole is centered at the origin. Results for a dipolar magnetic moment of a) $M=0$ and b) $M=475$ Am$^2$. The LAPD background magnetic field is represented by $B_0$.}%
    \label{fig:experiment}%
\end{figure}

 The experimental platform was developed on the LAPD facility at UCLA. In the experiments, a high-intensity laser was focused onto a solid target, releasing a fast-expanding plasma into the uniform, magnetized background plasma generated by the LAPD. A dipole magnet was inserted in the center of the background plasma. By measuring 2D planes of the magnetic field with motorized probes, the main characteristics of ion-scale magnetospheres were identified for different magnetic moments of the dipole~\cite{Schaeffer2022}. In Fig.~\ref{fig:experiment}, we observe the variation of the magnetic field $\Delta B_z\equiv B_z-B_{z,\text{ini}}$ over time, where $B_z$ and $B_{z,\text{ini}}$ are the total and initial magnetic fields, respectively, for the cases with a) no dipole and b) a moderate dipolar magnetic moment. These results were taken along the main plasma flow direction $y$, in the axes of symmetry $x=z=0$.%

\par In the case of Fig.~\ref{fig:experiment} a), there is no dipolar magnetic field present. As the driver flows against the magnetized background plasma, it expels the magnetic field in the upstream region, creating a magnetic cavity, while compressing it downstream, where the background is located. For the conditions of this experiment, the velocities of the magnetic cavity and magnetic compression are approximately constant, with values of 135 km/s and 380 km/s, respectively. We also observe that, when the driver runs out of energy, the cavity stops expanding. These results are consistent with previous LAPD experiments~\cite{VanZeeland2004,Collette2010,Niemann2013,Clark2013,Schaeffer2015,Schaeffer2018,Bonde2018II}. 

\par In Fig.~\ref{fig:experiment} b), we have the case with a moderate dipolar magnetic field. During the initial times of the experiment, the dipole can be neglected, and we observe the same features as in Fig.~\ref{fig:experiment} a). In particular, we observe the magnetic cavity and compression moving approximately at the same velocities as in the case without a dipole. When the dipolar magnetic field becomes strong enough, the plasmas are reflected, and we observe a reflection of the magnetic compression. These features were observed for multiple magnetic moments.%

\par
\begin{figure}[!h]
    \centering
    \includegraphics[width=0.95\columnwidth]{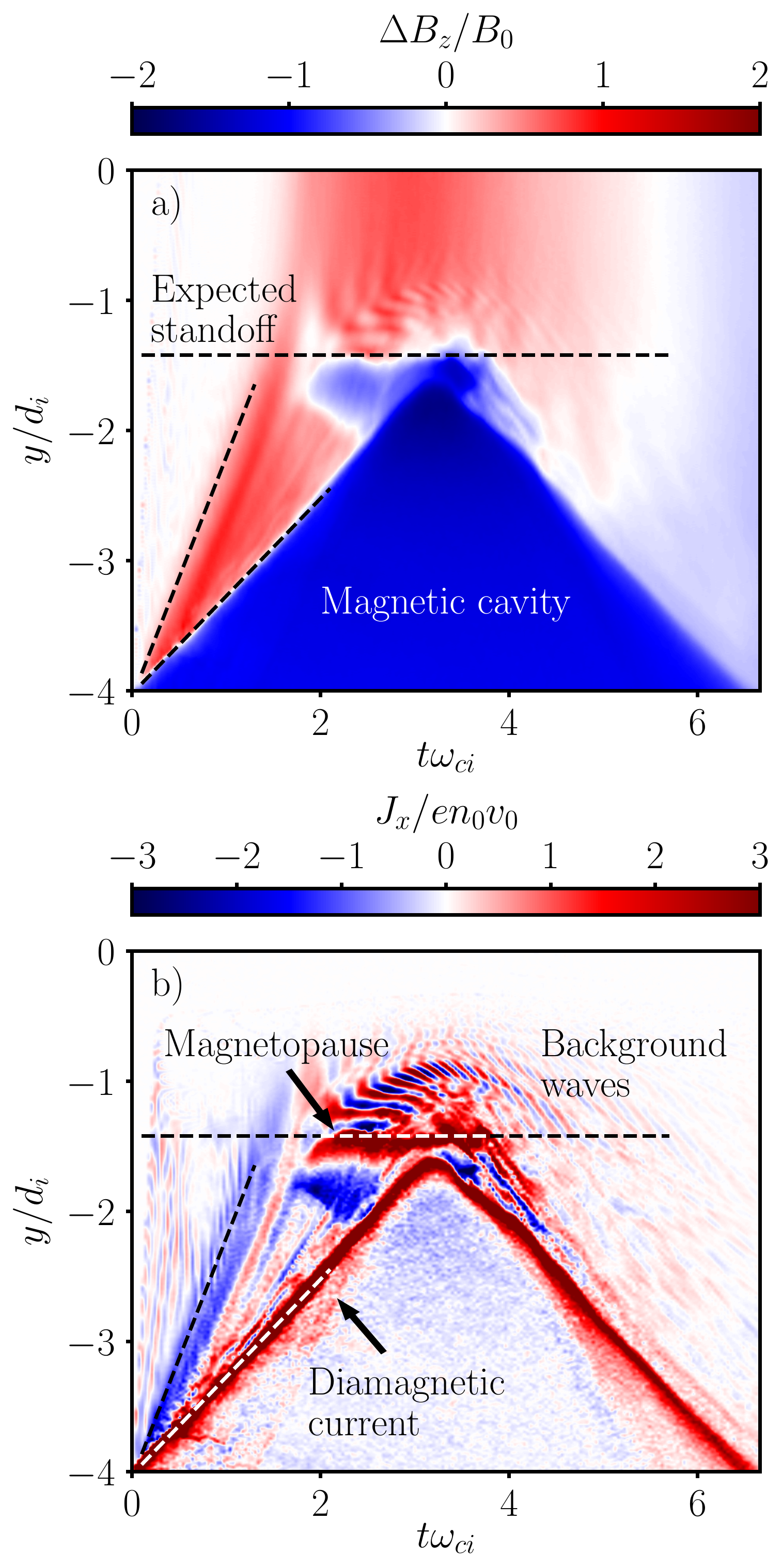}
    \vspace{-2ex}
    \caption{Temporal evolution of (a) the variation of the magnetic field $\Delta B_z$ and (b) the current density $J_x$ at the symmetry axis $x=0$, in the ``dayside'' region, for a simulation of laboratory ion-scale magnetospheres in similar conditions to the experiment in Fig.~\ref{fig:experiment} b).}%
    \label{fig:simulation}%
\end{figure}

 To explain the features in the experiments, and understand their dependency with the parameters, we performed multiple 2D PIC simulations of laboratory ion-scale magnetospheres, with a simplified setup of the experiments~\cite{Cruz2022}. These simulations considered a uniform driver plasma flowing against a uniform, magnetized background plasma with a dipolar magnetic field located in the center. In Fig.~\ref{fig:simulation}, we observe the variation of the magnetic field $\Delta B_z$ and of the current density $J_x$ at the axis of symmetry $x=0$, for a simulation in similar conditions to the experiment in Fig.~\ref{fig:experiment} b).%

\par Similar to the experimental plot in Fig.~\ref{fig:experiment} b), Fig.~\ref{fig:simulation} a) shows the formation of a magnetic cavity upstream and a magnetic compression downstream, in the initial times of the simulation, where the dipole is negligible. The plasmas continue to approach the dipole until the magnetic field is strong enough to reflect the magnetic compression. The simulations showed that the observation of this reflection depends on the size of the driver.%

\par Fig.~\ref{fig:simulation} b) shows the current density $J_x$ of the simulation. We observe two main current structures, namely, the diamagnetic current that supports the magnetic cavity in the upstream region, and the magnetopause current. The standoff locations for these currents can be estimated by the pressure balance between the ram pressure exerted by the plasmas and the magnetic field pressure 
\begin{equation}
    nm_iv_0^2 = \frac{B_z^2}{8\pi} \ ,
    \label{eq:pressure-mag}
\end{equation}
\noindent where $n$, $m_i$, and $v_0$ represent the ion density, mass, and fluid velocity of the plasma. We observed that the magnetopause current is more easily identified for low magnetic moments and that it is supported by the background and driver plasmas with some time dependence. These results were also consistent with the experiments.%

\par Since the plasmas interact with each other before being stopped by the dipole at the standoff distance, the features observed in the experimental and numerical studies of laboratory ion-scale magnetospheres are dependent on the coupling strength between the two plasmas. The densities affect this coupling~\cite{Hewett2011}, leading to faster magnetic cavities for higher driver densities, which results in different magnetospheric features~\cite{Cruz2022}. Additionally, even without the dipole, the driver continuously loses energy to the background and eventually stops expanding, as seen in Fig.~\ref{fig:experiment} a). To observe the reflection of the magnetic compression, we need to ensure that the driver has enough energy to reach the standoff distance, which depends on the coupling~\cite{Ripin1993,Winske2007,Cruz2022}.%

\par To better understand and design experiments with laboratory ion-scale magnetospheres, we then need to accurately describe the coupling of the system. In Secs.~\ref{sec:simulations} to~\ref{sec:comparison}, we present a model capable of describing this coupling, for a range of parameters relevant to current experimental facilities.%

\section{\label{sec:simulations} PIC simulations}

\subsection{\label{sub:configuration} Configuration of the simulations}

\par To study the coupling between an unmagnetized driver plasma and a magnetized background plasma, we performed multiple 1D simulations with OSIRIS, a massively parallel and fully relativistic PIC code~\cite{Fonseca2002,Fonseca2013}. With PIC simulations, we can accurately resolve the plasma kinetic scales of these systems.%

\par The simulations consist of a 25 $d_i$ length region with open boundary conditions at $y=-5\ d_i$ and $y=20\ d_i$, where $d_i=c/\omega_{pi}=\sqrt{m_{i,0}c/4\pi n_0e^2}$ is the ion skin depth of the background plasma, with $c$ the speed of light in vacuum, $\omega_{pi}$ the ion plasma frequency, $e$ the electron charge, and $m_{i,0}$ and $n_0$ the ion mass and the density of the background plasma, respectively. To resolve the dynamics of the electron kinetic scales, we used 10 grid cells per electron skin depth $d_e=d_i\sqrt{m_e/m_{i,0}}$, where $m_e$ is the electron mass.%

\par For the driver plasma, the simulations consider an idealized and simplified configuration when compared to typically laser-produced driver plasmas in the laboratory~\cite{Zakharov2003,Schaeffer2018}. The driver has an initial fluid velocity $\mathbf{v_0}=v_0{\ \mathbf{\hat{y}}}$, a uniform density $n_d$, and a length $L_y=5\ d_i$. The driver is initially located between $y=-5\ d_i$ and $y=0$, and it is composed of electrons and a single species of ions with mass $m_{i,d}$. Equivalently, the background plasma has a density of $n_0$ and a length of $L_B=20\ d_i$. It is located between $y=0$ and $y=20\ d_i$, and it is also composed of electrons and a single species of ions with mass $m_{i,0}$. Unlike the driver, the background plasma is magnetized by an internal and uniform magnetic field $\mathbf{B_0}=B_0\ \mathbf{\hat{z}}$. The magnitude $B_0$ is calculated from the Alfvénic Mach number, defined as $M_A\equiv v_0/v_A=v_0\sqrt{4\pi n_0m_{i,0}}/B_0$, where $v_A$ is the Alfvén velocity. Both plasmas have 200 particles per cell per species and ions with charge $q_i=e$.%

\par We consider electron thermal velocities of $v_{the}=0.1\ v_0$, with $v_{the,x}=v_{the,y}=v_{the,z}=v_{the} / \sqrt{3}$, and that the ions and electrons are initially in thermal equilibrium. Since the most relevant dynamics of the simulations occur at the ion kinetic scales, the spatial scales are normalized to $d_i$ and the time scales to the ion cyclotron gyroperiod of the background plasma $\omega_{ci}^{-1}=m_{i,0}c/eB_0$.%

\par The simulations consider colder plasmas, lower ion mass ratios $m_{i,0}/m_e = 100$, and faster fluid velocities $v_0=0.1\ c$ than expected in experiments and most space and astrophysical scenarios. These approximations reduce the computational resources necessary to perform the simulations, allow extended scans over different parameters, and simplify our analysis. The chosen ion-to-electron mass ratio in the simulations is high enough to ensure sufficient separation between electron and ion spatial and temporal scales. 

\par Additionally, $n_0$ is the independent variable of OSIRIS. We ensure that $v_0$ is low enough to neglect relativistic effects on the system. By using proper space and time scales ($d_i$, $\omega_{ci}^{-1}$), we expect the main properties of the system to scale with the main dimensionless parameters ($M_A$, $n_d/n_0$, and $m_{i,d}/m_{i,0}$). Therefore, the simplifications considered in the simulations should not affect the main results.%

\par The main parameter scans presented in this paper consider $0.2 \leq n_d/n_0 \leq 10$, $1 \leq m_{i,d}/m_{i,0} \leq 9$ and low Alfvénic Mach numbers such that $0.5\leq M_A \leq 1.5$. Later simulations also consider $2\leq M_A \leq 10$ and longer plasmas with $L_y=120\ d_i$ and $L_B=300\ d_i$. During the parameter scans, we keep the background parameters $n_0$ and $m_{i,0}$ unchanged, and instead change the driver parameters $n_d$ and $m_{i,d}$, and the Mach number $M_A$. For these parameters, and for the density profiles considered, the lengths and times of the simulations are long enough to observe a quasi-steady-state regime of the system for a sufficient amount of time~\cite{Clark1973} and a strong coupling regime between the two plasmas~\cite{Schaeffer2015}.%

\subsection{\label{sub:dynamics} Basic system dynamics}

\par
\begin{figure*}[!htp]
    \centering
    \includegraphics[width=0.88\linewidth]{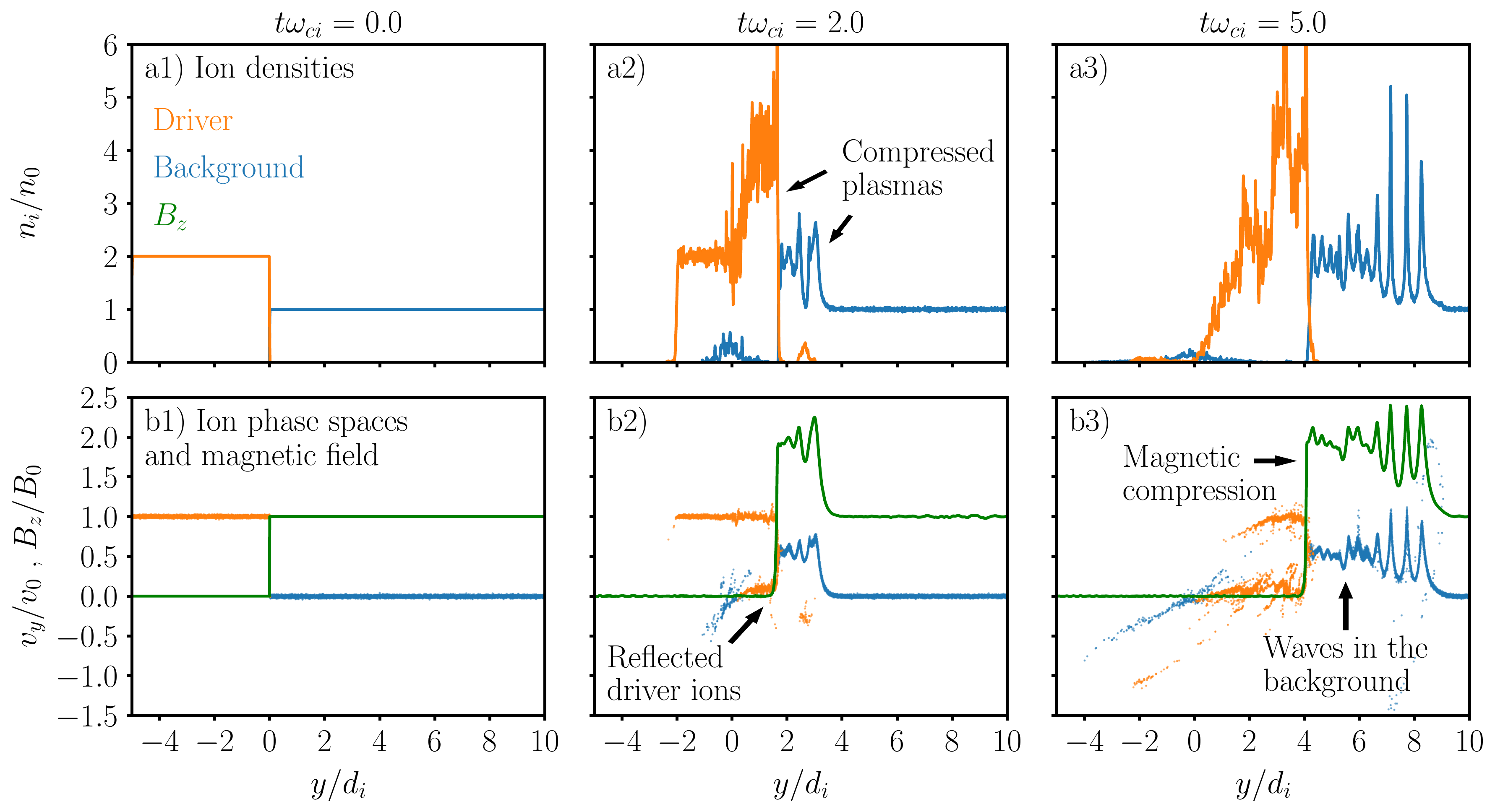}
    \caption{\label{fig:basic} a) Ion densities $n_i$ and b) $y$ component of the ion velocities $v_y$, for the driver (orange) and background (blue) plasmas. The green line shows the magnetic field $B_z$. Columns 1--3 correspond to three different times. The simulation considers $n_d/n_0=2$, $m_{i,d}/m_{i,0}=1$, and $M_A=1.5$. [Associated dataset available at https://zenodo.org/record/7485077 (Ref.~\onlinecite{Zenodo}).]}%
\end{figure*}

 Fig.~\ref{fig:basic} illustrates the basic temporal evolution of the system and shows the ion densities $n_i$ of the driver and background plasmas, the ion phase spaces, and the magnetic field $B_z$, for three different times. The initial setup of the simulations is shown in Figs.~\ref{fig:basic} a1) and b1). The simulation represented considers $n_d/n_0=2$, $m_{i,d}/m_{i,0} = 1$, and $M_A=1.5$.%

\par We see in Figs.~\ref{fig:basic} a1--3) that, as the driver flows to the right, it pushes the background plasma and the magnetic field with it, leading to a relocation of the interface between the two plasmas and creating two high-density regions on both sides of the interface. During this process, the driver ions are mostly confined in the upstream region relative to the plasma flow, while the background ions are mostly confined downstream, with the exception of small amounts of the driver and background ions that entered the opposite regions in the initial times of the simulation~\cite{Takezaki2016,Takezaki2021}.%

\par The magnetic field rapidly increases in the transition from the unmagnetized driver to the magnetized background, as we can see in Figs.~\ref{fig:basic} b1--3). Due to the large mass discrepancies between the electrons and the ions, the transition occurs over a length scale larger than the local electron gyroradius but much smaller than the local ion gyroradius, and so, in the interface region between the two plasmas, the driver electrons are effectively magnetized while the driver ions are unmagnetized. The space-charge separation creates a negative electric field in the $y$ direction that reflects the driver ions back upstream with a new velocity $v_1<v_0$, and causes an electron $\mathbf{E}\times \mathbf{B}$ drift, which leads to a diamagnetic current. This current expels the magnetic field upstream, creating a magnetic cavity with no magnetic field, while the field is compressed downstream~\cite{Ripin1993,VanZeeland2004,Collette2010}. A more detailed study of the electric fields of the transition region is presented in Appendix~\ref{sec:appendix}.%

\par The energy and momentum lost by the driver plasma during this process are transferred to the background region. The initially stationary background ions are accelerated and the magnetic field is compressed. From this interaction, the bulk of the driver, the compressed background, and the interface between the plasmas travel to the right through the region initially occupied by the background plasma. Additionally, during this process, multiple waves and instabilities form in the background region, as we see in Fig.~\ref{fig:basic} b3). We also observe that the size of the perturbed background region increases over time.%

\par In Fig.~\ref{fig:basic}, the magnetic field and the plasma densities of the system are not constant, however, some average quantities of the system do not change significantly over time. Such examples include the average velocity of the accelerated background ions and the average of the compressed magnetic field. We could, therefore, consider that the system achieves a quasi-steady-state regime and that it can be represented by its average properties, as we show in Sec.~\ref{sec:derivation}.%

\subsection{\label{sub:streak} Magnetic field diagnostics}%

\par To comprehend the dynamics of these systems, it is important to investigate the evolution of the magnetic field $B_z$ since it can be used to determine the motion of the particles and the electric fields. Additionally, magnetic field diagnostics are widely used for fast-driven plasmas in laboratory experiments~\cite{VanZeeland2004,Schaeffer2018,Schaeffer2022}. From Amp\`{e}re's law, the $x$ component of the current density is $J_x \approx (c/4\pi)\cdot \partial B_z/\partial y$, and so, we can use $J_x$ to investigate changes in the magnetic field.%

\par
\begin{figure*}[!htp]
    \centering
    \includegraphics[width=\linewidth]{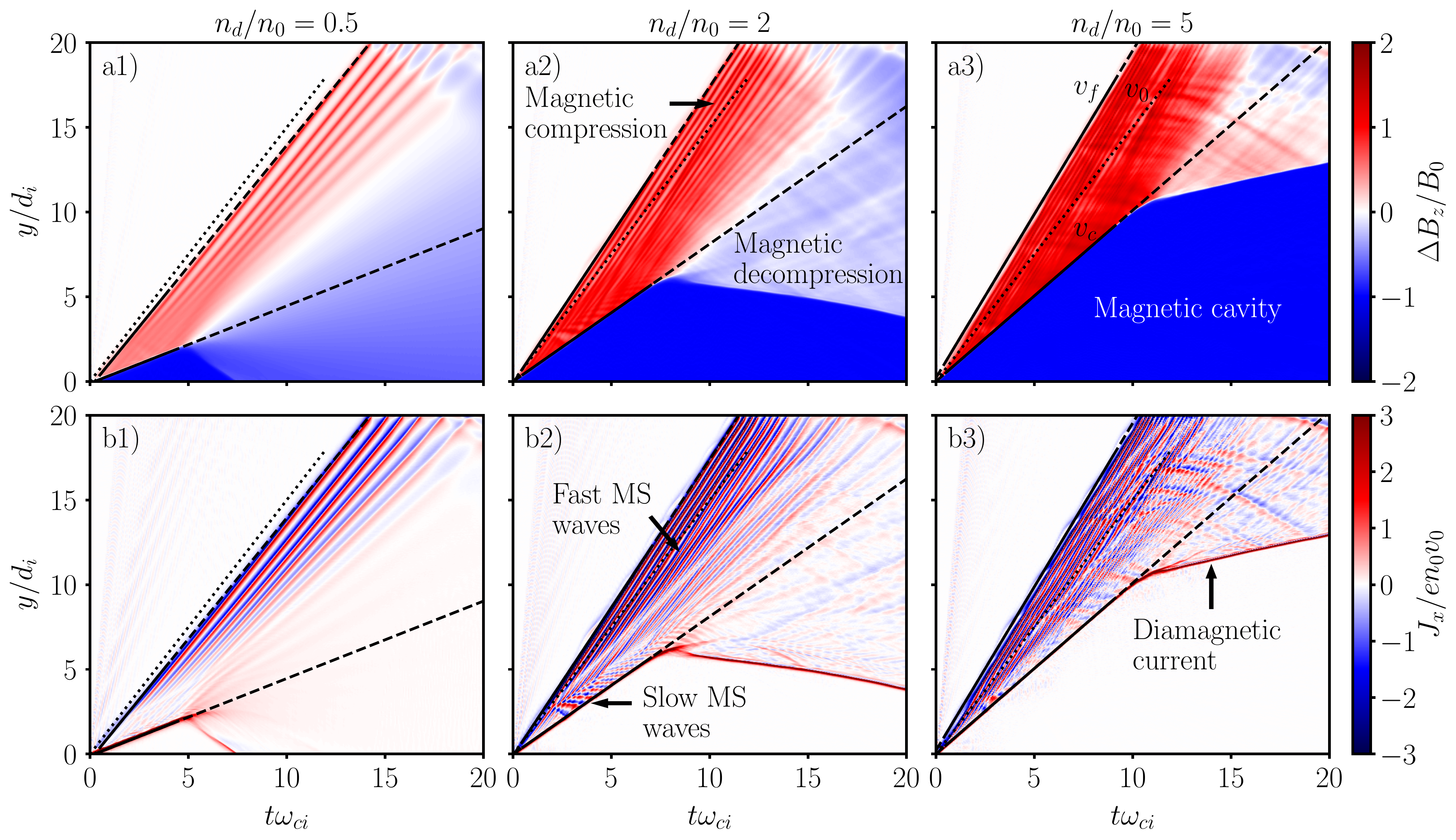}
    \vspace{-4ex}
    \caption{\label{fig:coupling-streak-plots-high} Temporal evolution of a) the variation of the magnetic field $\Delta B_z$ and b) the current density $J_x$ for three different simulations with 1) $n_d/n_0=0.5$, 2) $n_d/n_0=2$, and 3) $n_d/n_0=5$, with $m_{i,d}/m_{i,0}=1$ and $M_A=1.5$. The dotted line has a slope equal to $v_0$, and the solid and dashed lines have slopes equal to the coupling velocity $v_c$ and the front velocity $v_f$. Frames a2) and b2) correspond to the simulation shown in Fig.~\ref{fig:basic}. We observe magnetosonic (MS) waves in the background in all frames.}%
\end{figure*}

 To illustrate these two important quantities, we show in Fig.~\ref{fig:coupling-streak-plots-high} the temporal evolution of a) the variation of the magnetic field $\Delta B_z\equiv B_z-B_{z,\textrm{ini}}$, where $B_{z,\textrm{ini}}$ is the initial magnetic field, and b) the current density $J_x$. To understand how the system depends on the initial parameters, these diagnostics are shown for three driver densities: 1) $n_d/n_0=0.5$, 2) $n_d/n_0=2$, and 3) $n_d/n_0=5$, with $m_{i,d}/m_{i,0}=1$ and $M_A=1.5$.%

\par In Fig.~\ref{fig:coupling-streak-plots-high} we observe the same main structures for the three driver densities. Similarly to the experiment in Fig.~\ref{fig:experiment} a), and as discussed in Sec.~\ref{sub:dynamics}, while the driver flows against the background, it expels the magnetic field, creating a magnetic cavity with no magnetic field, as observed in Figs.~\ref{fig:coupling-streak-plots-high} a1--3). This magnetic cavity expands over time, and its maximum extent increases with the driver density. This is expected since the energy and the pressure exerted by the driver increase with its density, improving the coupling between the plasmas~\cite{Hewett2011}. The velocity at which the magnetic cavity travels through the background is designated by coupling velocity $v_c$. This velocity is always lower than $v_0$ and increases with the driver density, as shown in Fig.~\ref{fig:coupling-streak-plots-high}.%

\par After all the driver ions with velocity $v_0$ are reflected, the driver may not have enough energy to push the background any further, leading to the reflection of the magnetic cavity. This happens for $n_d/n_0 = 0.5$ and $n_d/n_0 = 2$ (at $t\omega_{ci} \approx 5$ and $t\omega_{ci} \approx 8$, respectively). For $n_d/n_0 = 5$, the reflected driver ions continue pushing the magnetic cavity through the background region, although at a lower velocity. We consider the stopping distance $L_{\textrm{stop}}$ the distance that the magnetic cavity travels before the driver ions with velocity $v_0$ are fully reflected, \textit{i.e.}, during the main interaction of the system.%

\par In Figs.~\ref{fig:coupling-streak-plots-high} a1--3), we also observe the magnetic compression in the downstream region, where the background is located. While the compressed magnetic field is not constant, its average does not change significantly over time during the main interaction of the plasmas. The average ratio of compressed to the initial magnetic field is designated by compression ratio $\alpha$. Additionally, the extent of the background plasma with compressed magnetic field increases over time. The velocity at which the magnetic compression travels through the unperturbed background plasma is designated by front velocity $v_f$. Fig.~\ref{fig:coupling-streak-plots-high} shows that $v_f$ also increases with the driver density. After the driver's reflection, the compressed plasma continues to move through the unperturbed background.%

\par Figs.~\ref{fig:coupling-streak-plots-high} b1--3) show the current densities for the different driver densities. The sudden increase in magnetic field from the magnetic cavity to the magnetized background plasma is supported by the diamagnetic current~\cite{Collette2010}. In the background region, we also observe multiple current structures associated with the fast and slow magnetosonic (MS) waves that form in the background plasma~\cite{Everson2016,Schaeffer2018}.%

\section{\label{sec:derivation} Coupling parameters}

\par
\begin{figure}[!h]
    \centering
    \begin{tikzpicture}
        \definecolor{mygreen}{HTML}{2CA02C}
        \definecolor{myorange}{HTML}{FF7F0E}
        \definecolor{myblue}{HTML}{1F77B4}
        
        \draw[thick,->] (-0.3,0) -- (6,0) node[anchor=north east] {$y$};
        \draw[thick,->] (0,-0.35) -- (0,2.9) node[anchor=north east] {$B_z$, $v_y$};
        \draw (-2.5pt,1.1cm) -- (2.5pt,1.1cm) node[anchor=east, auto, xshift=-0.1cm] {$B_0$};
        \draw (-2.5pt,2.2cm) -- (2.5pt,2.2cm) node[anchor=east, auto, xshift=-0.1cm] {$\alpha B_0$};
        \draw (-2.5pt,0cm) -- (2.5pt,0cm) node[anchor=north east, auto, xshift=-0.1cm] {$0$};
        
        \draw[line width=2pt,mygreen] (0,0) -- (2,0) -- (2,2.2) -- (4,2.2) -- (4,1.1) -- (5.5,1.1) ;
        \draw[thick,->] (2,1.7) -- (2.7,1.7) node[above, midway] {$\mathbf{v_c}$};
        \draw[thick,->] (4,1.7) -- (5.0,1.7) node[above, midway] {$\mathbf{v_f}$};
        
        \draw[dashed] (0,2.2) -- (2,2.2) ;
        \draw[dashed] (0,1.1) -- (4,1.1) ;
        \draw[dashed] (2,-0.35) -- (2,2.5) node[anchor=south] {$A$};
        \draw[dashed] (4,-0.35) -- (4,2.5) node[anchor=south] {$B$};
        
        \draw [decorate, decoration = {calligraphic brace, mirror,amplitude=5pt}, line width=1.5pt] (0.1,-0.25) --  (1.9,-0.25) node[below,midway,yshift=-5pt]{ \begin{tabular}{c} Magnetic\\ cavity \end{tabular}};
        \draw [decorate, decoration = {calligraphic brace, mirror,amplitude=5pt}, line width=1.5pt] (2.1,-0.25) --  (3.9,-0.25) node[below,midway,yshift=-5pt]{\begin{tabular}{c} Magnetic\\ compression \end{tabular}};
        \draw [decorate, decoration = {calligraphic brace, mirror,amplitude=5pt}, line width=1.5pt] (4.1,-0.25) --  (5.5,-0.25) node[below,midway,yshift=-5pt,xshift=0pt]{ \begin{tabular}{c} Unperturbed\\ background \end{tabular}};
        
        \draw[color=myorange] plot [only marks, mark=*,mark size=0.3, domain=0.05:1.95, samples=150] (\x,{1.3+rnd*0.15});
        \draw[thick,->] (0.4,1.7) -- (1.4,1.7) node[above, midway] {$\mathbf{v_0}$};
        \draw[color=myorange] plot [only marks, mark=*,mark size=0.3, domain=1.3:1.95, samples=70] (\x,{0.1+rnd*0.15});
        \draw[thick,->] (1.47,0.4) -- (1.87,0.4) node[above, midway] {$\mathbf{v_1}$};
        \draw[color=myorange] plot [only marks, mark=*,mark size=0.3, domain=0.1:1.45, samples=70] ({1.96+rnd*0.08},\x);
        \draw[color=myblue] plot [only marks, mark=*,mark size=0.3, domain=-0.1:0.85, samples=60] ({1.96+rnd*0.08},\x);
        \draw[color=myblue] plot [only marks, mark=*,mark size=0.3, domain=2.05:3.95, samples=150] (\x,{0.7+rnd*0.15});
        \draw[thick,->] (2.7,0.2) -- (3.4,0.2) node[above, midway] {$\mathbf{v_m}$};
        \draw[color=myblue] plot [only marks, mark=*,mark size=0.3, domain=4.05:5.5, samples=200] (\x,{-0.1+rnd*0.15});
        \draw[color=myblue] plot [only marks, mark=*,mark size=0.3, domain=-0.1:0.85, samples=60] ({3.96+rnd*0.08},\x);

    \end{tikzpicture}
    \caption{\label{fig:model} Simplified representation of the interaction between the flowing driver (orange) with the background (blue) and the magnetic field profile (green). All quantities represented are measured in the lab frame. The dots illustrate the plasma particles' velocities. This model considers three regions: the magnetic cavity, the magnetic compression, and the unperturbed background. These regions are separated by the discontinuities $A$ and $B$.}%
\end{figure}
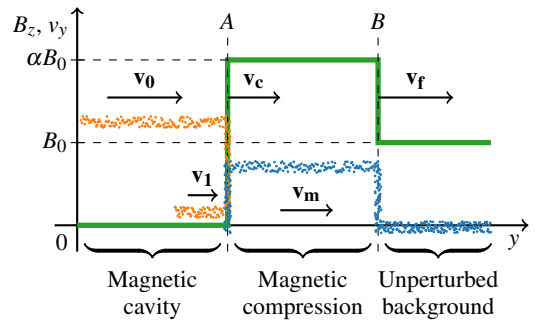

 As discussed in Sec.~\ref{sub:dynamics}, for uniform densities, and if the plasmas are long enough, the system reaches a quasi-steady-state regime, where some average quantities do not change significantly over time. Under these conditions and assuming the MHD formalism, we can describe the system by three different regions with different magnetic and kinetic properties, as shown in Fig.~\ref{fig:model}.%

\par The first region in Fig.~\ref{fig:model} refers to the magnetic cavity, where $B_z\approx0$. The driver plasma is located in this upstream region, and the ion motion of the driver can be described by two ion populations with velocities $\mathbf{v_0}$ and $\mathbf{v_1}$. The second region refers to the magnetic compression, where the average magnetic field is $\alpha B_0$. In this region, the background ions accelerated by the interaction with the driver plasma have an average velocity of $\mathbf{v_m}$. Finally, the third region refers to the background region that remains unperturbed by the interaction of the two plasmas. Here, the magnetic field is $B_0$ and the background ions have no flow velocity. These three regions are visible in Fig.~\ref{fig:coupling-streak-plots-high}.%

\par The regions are separated by two discontinuities where the plasma properties change abruptly. Discontinuities $A$ and $B$ move through the simulation box with velocity $\mathbf{v_c}$ and $\mathbf{v_f}$, respectively. Some physical quantities must be conserved from the upstream to the downstream side of the discontinuities. With these conservation laws, we can derive analytical expressions for the coupling parameters $\alpha$, $v_c$, and $v_f$.%

\subsection{\label{sub:jump} Jump conditions at the discontinuities}

\par To derive expressions for the coupling parameters, we consider the system description in Fig.~\ref{fig:model}. In the reference frame of the discontinuities, and for a quasi-steady-state regime, the time derivatives can be dropped in the MHD and Maxwell's equations.%

\par In the upstream and downstream sides of each discontinuity, some physical quantities must be conserved, which leads to the Rankine-Hugoniot (R-H) jump conditions. For magnetic fields perpendicular to the plasma flows, these conditions lead to~\cite{Kivelson1995,Balogh2013,Thompson2013,Vink2020}
\begin{align}
    & \left[vB\right] = 0 \label{eq:ohm_faraday_laws-rh} \\
    & \left[n_iv\right] = 0 \label{eq:mass_conservation-rh} \\
    & \left[n_im_iv^2+p+\frac{B^2}{8\pi}\right] =0 \label{eq:momentum_conservation-rh} \\
    & \left[n_im_{i}v\left(\frac{1}{2}v^2+u+\frac{p+B^2/8\pi}{n_im_{i}}\right)\right] = 0 \ . \label{eq:energy_conservation-rh}
\end{align}
In the previous equations, $[X]\equiv X_1-X_2$ stands for the difference of values of a quantity $X$ in the upstream and downstream regions, while $n_i$, $m_i$, and $v$ stand for the density, mass, and velocity of the ions, respectively, $p$ for the electron pressure, $u$ for the internal energy per unit mass, and $B$ for the magnetic field. Eq.~\eqref{eq:ohm_faraday_laws-rh} corresponds to Ohm's law, and Eqs.~\eqref{eq:mass_conservation-rh} to~\eqref{eq:energy_conservation-rh} to the mass, momentum, and energy conservation laws. We can apply these equations to both discontinuities A and B in Fig.~\ref{fig:model}.%

\subsection{\label{sub:jumpA} R-H equations at discontinuity A}

\par From Eq.~\eqref{eq:ohm_faraday_laws-rh}, the quantity $v_yB_z$ must be conserved in the reference frame of discontinuity $A$. This leads to%
\begin{equation}
    (v_m-v_c)\alpha B_0 = 0 \iff v_m = v_c \ .
    \label{eq:vmvc}
\end{equation}
By also applying Eqs.~\eqref{eq:mass_conservation-rh} and~\eqref{eq:energy_conservation-rh} to the discontinuity, we obtain%
\begin{equation}
    \hspace{7ex} n_dm_{i,d}(v_0-v_c) + n_d'm_{i,d}(v_1-v_c) = 0 \label{eq:mass-cons}
\end{equation}
\vspace{-4ex}
\begin{align}
    \frac{1}{2}(v_0-v_c)^2-\frac{1}{2}(v_1-v_c)^2&+u_d-u_d'\nonumber \\ &+\frac{p_d}{n_dm_{i,d}}-\frac{p'_d}{n_d'm_{i,d}}=0 \ , \label{eq:ener-cons}
\end{align}
\noindent where $n_d'$, $p_d'$, and $u_d'$ are the density, pressure, and internal energy density of the driver population with velocity $v_1$. If we assume similar thermal properties for the two driver populations, the enthalpy terms $w=u+p/nm$ in Eq.~\eqref{eq:ener-cons} cancel out. Eqs.~\eqref{eq:mass-cons} and~\eqref{eq:ener-cons} then lead to $n_d'=n_d$ and $v_c=(v_0+v_1)/2$, as expected.%
\par Finally, from Eq.~\eqref{eq:momentum_conservation-rh}, the momentum must also be conserved, which leads to%
\begin{align}
    & 2n_dm_{i,d}(v_0-v_c)^2 + p_d+p_d' = p_2+\frac{(\alpha B_0)^2}{8\pi} \nonumber \\  
    \iff &\left(\frac{M_A}{R_n}\right)^2\left(1-\frac{v_c}{v_0}\right)^2=\alpha^2+\beta_2-\beta_d \ ,
    \label{eq:pressure-balance}
\end{align}
\noindent where the factor of 2 considers both populations of driver ions with $v_0$ and $v_1$ velocities, $p_2$ corresponds to the electron pressure in the compressed background region, and $\beta_d\equiv8\pi(p_d+p_d')/B_0^2$, $\beta_2\equiv8\pi p_2/B_0^2$, and%
\begin{equation}
    R_n \equiv \frac{1}{2}\left(\frac{n_0}{n_{d}}\frac{m_{i,0}}{m_{i,d}}\right)^\frac{1}{2} \ .
    \label{eq:Rn}
\end{equation}
To determine the other equations for the coupling parameters, we now apply Eqs.~\eqref{eq:ohm_faraday_laws-rh} to~\eqref{eq:energy_conservation-rh} to discontinuity $B$.%

\subsection{\label{sub:jumpB} R-H equations at discontinuity B}

\par By applying Eqs.~\eqref{eq:ohm_faraday_laws-rh} and~\eqref{eq:mass_conservation-rh} to the reference frame of discontinuity $B$, we obtain%
\begin{align}
    (v_m-v_f) \alpha B_0 = -v_f B_0 &\iff \alpha = \frac{v_f}{v_f-v_c} \label{eq:alphavcvf} \\
    (v_m-v_f)n_2 = -v_f n_0 &\iff \alpha = \frac{n_2}{n_0} \ , \label{eq:ndn0alpha}
\end{align}
where $n_2$ refers to the density of the compressed background plasma. Eqs.~\eqref{eq:alphavcvf} and~\eqref{eq:ndn0alpha} show that the background density and magnetic field increase by the same ratio~\cite{Schaeffer2015,Everson2016}. Eq.~\eqref{eq:alphavcvf} also shows a dependency between the three coupling parameters $\alpha$, $v_c$, and $v_f$.%

\par Using Eq.~\eqref{eq:momentum_conservation-rh}, we have from the momentum conservation in $B$ that%
\begin{align}
    &\alpha n_0m_{i,0}\left(\frac{v_f}{\alpha}\right)^2+p_2+\frac{(\alpha B_0)^2}{8\pi} =  n_0m_{i,0}v_f^2+p_1+\frac{B_0^2}{8\pi}\nonumber \\ 
    \iff &\beta_2=\beta_1+(1-\alpha^2)+2M_B^2\left(1-\frac{1}{\alpha}\right) \ , \label{eq:beta2}
\end{align}
where $p_1$ is the initial background electron pressure, $\beta_1\equiv 8\pi p_1/B_0^2$, and $M_B\equiv v_f/v_A$. With Eq.~\eqref{eq:energy_conservation-rh} and using $u=p/(\gamma-1)n_im_i$, where $\gamma$ is the adiabatic index, we obtain%
\begin{equation}
    M_B^2\left(1-\frac{1}{\alpha^2}\right)+\frac{\gamma}{(\gamma-1)}\left(\beta_1-\frac{\beta_2}{\alpha}\right)+2(1-\alpha)=0 \ . \label{eq:energy2}
\end{equation}
Finally, Eqs.~\eqref{eq:beta2} and~\eqref{eq:energy2} allow us to derive an expression for the compression ratio~\cite{Cairns1994,Kivelson1995,Everson2016,Vink2020}
\begin{widetext}
\begin{equation}
    \alpha = \frac{2(\gamma+1)}{\gamma-1}\left[1+\frac{\gamma(1+\beta_1)}{\gamma-1}M_B^{-2} +\left(\left[1+\frac{\gamma(1+\beta_1)}{\gamma-1}M_B^{-2}\right]^2+4\frac{(1+\gamma)(2-\gamma)}{(\gamma-1)^2}M_B^{-2}\right)^{1/2}\right]^{-1} \ .
    \label{eq:large}
\end{equation}
\end{widetext}
Eq.~\eqref{eq:large} shows an expression for $\alpha$ as a function of the velocity of the discontinuity. By solving Eqs.~\eqref{eq:pressure-balance}, \eqref{eq:alphavcvf}, and~\eqref{eq:large} numerically, we can determine the coupling parameters $\alpha$, $v_c$, and $v_f$ with only the initial conditions of the system.

\subsection{\label{sub:parameters} Solutions for low Mach numbers}

\par With Eqs.~\eqref{eq:pressure-balance} and~\eqref{eq:large}, and by using Eq.~\eqref{eq:beta2} to replace $\beta_2$ in Eq.~\eqref{eq:alphavcvf}, we obtain a set of three equations that relate the coupling parameters $\alpha$, $v_c$, and $v_f$ with each other and with the initial parameters of the system. By solving these equations numerically, we can calculate these parameters for given initial conditions. In some regimes, however, it is possible to also obtain analytical expressions for the coupling parameters.%

\par From Eq.~\eqref{eq:large}, it can be shown that for cold plasmas, such that $\beta_1,\beta_d\ll 1$, and low Mach numbers, we can consider $\alpha\approx M_B\equiv v_f/v_A$ (see Appendix~\ref{app:energy}). Replacing Eq.~\eqref{eq:large} by this approximation, and using Eqs.~\eqref{eq:pressure-balance} and~\eqref{eq:alphavcvf}, we have that $\alpha$, $v_c$, and $v_f$ are given by
\begin{align}
    \alpha &= \frac{1+M_A}{1+R_n}
    \label{eq:alpha} \\
    \frac{v_c}{v_0} &= \frac{1}{M_A}\frac{M_A-R_n}{1+R_n}
    \label{eq:vcv0} \\
    \frac{v_f}{v_0} &= \frac{1}{M_A}\frac{1+M_A}{1+R_n} \ .
    \label{eq:vfv0}
\end{align}
These expressions depend only on the initial parameters of the system. We stress that $\alpha$, $v_c$, and $v_f$ can be directly measured from the magnetic diagnostics of the experiments and of the simulations, and therefore, they can be used to evaluate the coupling between the plasmas and to estimate uncertain initial conditions of the system.

\par The equations shown here are only valid for the main interaction of the system, \textit{i.e.}, before all the driver ions get reflected. After the main interaction, these reflected ions may still have enough energy to continue pushing the background forward, creating a second interaction between the driver and the background, but with different parameters.%

\subsection{\label{sec:parameters-high} Solutions for high Mach numbers}

\par We can also obtain analytical expressions for the coupling parameters for cold plasmas and high Mach numbers, such that $\beta_1,\beta_d\ll 1$ and $M_A\gg1$. Under this conditions, Eqs.~\eqref{eq:pressure-balance}, \eqref{eq:alphavcvf}, and~\eqref{eq:large} lead to
\begin{gather}
    \alpha = \frac{\gamma+1}{\gamma-1} \label{eq:alpha-high} \\
    \frac{v_c}{v_0} = \frac{1}{1+R_n\sqrt{1+\gamma}} \label{eq:vcv0-high} \\
    \frac{v_f}{v_0} = \frac{1+\gamma}{2(1+R_n\sqrt{1+\gamma})} \label{eq:vfv0-high} \ .
\end{gather}
Eq.~\eqref{eq:alpha-high} represents the maximum value for the compression ratio, well-known in strong shock theory~\cite{Kivelson1995,Balogh2013,Vink2020}.%

\section{\label{sec:comparison} Comparison between the equations and the simulations}

\subsection{\label{sub:scan_density} Dependency on the driver density}

\par After deriving analytical expressions that describe the coupling between the driver and background plasmas, we now verify the validity of Eqs.~\eqref{eq:alpha} to~\eqref{eq:vfv0}. For the study of ion-scale magnetospheres, we are interested in the cases where $M_A\sim1$ and $n_d\sim n_0$. For these conditions, we can apply the solutions for low Mach numbers, represented in Sec.~\ref{sub:parameters}.%

\par
\begin{figure}[!h]
    \centering
    \includegraphics[width=0.95\columnwidth]{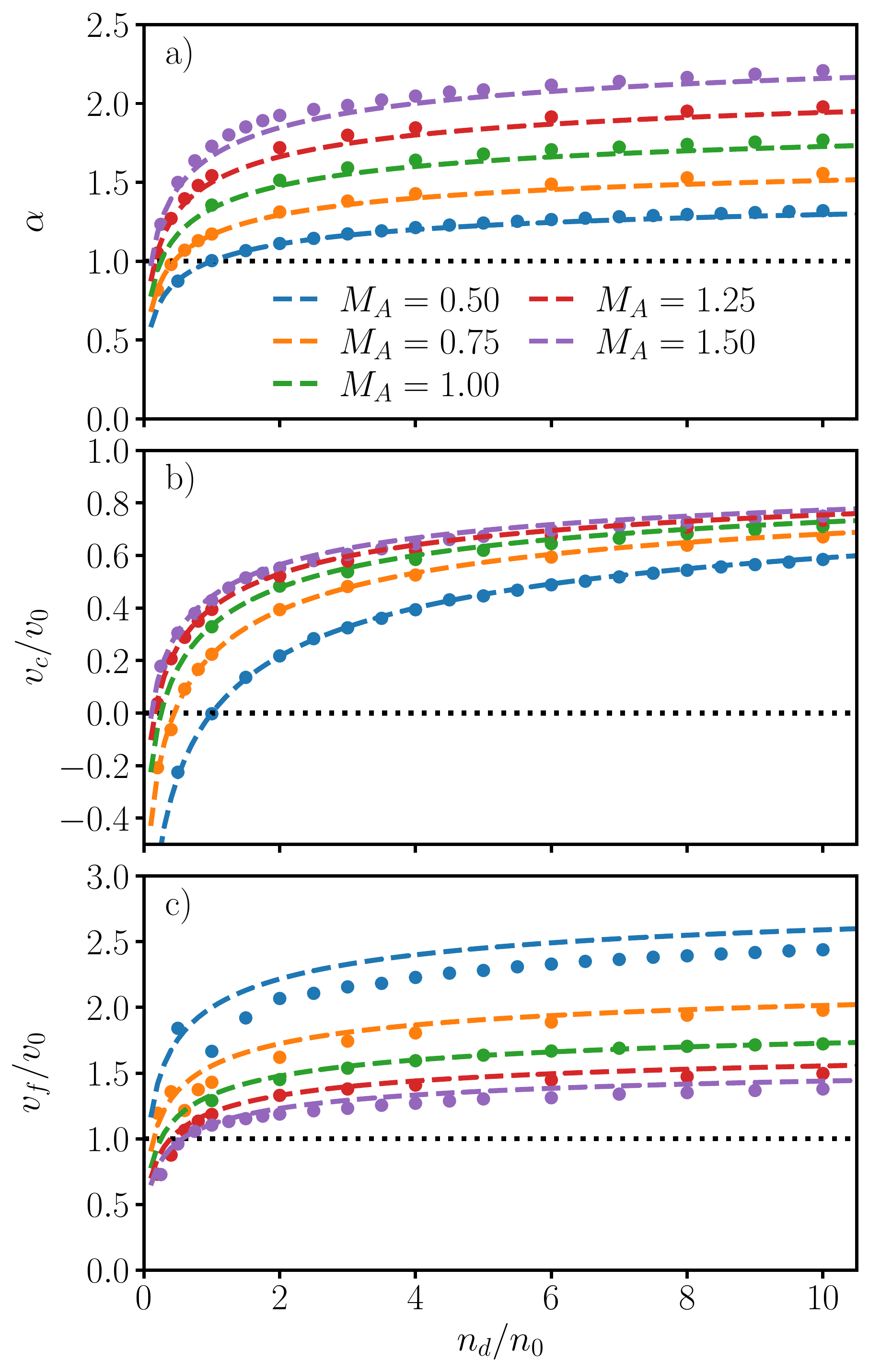}
    \vspace{-2ex}
    \caption{Comparison between the coupling parameters measured in the simulations for a) $\alpha$ with Eq.~\eqref{eq:alpha}, b) $v_c/v_0$ with Eq.~\eqref{eq:vcv0}, and c) $v_f/v_0$ with Eq.~\eqref{eq:vfv0}, for different $M_A$ and $n_d/n_0$ values. These simulations considered $m_{i,d}/m_{i,0}=1$. The coupling parameters measured in the simulations are represented by dots and the analytical expressions by dashed lines. [Associated dataset available at https://zenodo.org/record/7485077 (Ref.~\onlinecite{Zenodo}).]}%
    \label{fig:scan} 
\end{figure}

 Fig.~\ref{fig:scan} shows the measured values for the coupling parameter ratios $\alpha$, $v_c/v_0$, and $v_f/v_0$, for multiple simulations with different Alfvénic Mach numbers $M_A$ and driver-to-background density ratios $n_d/n_0$, with $m_{i,d}/m_{i,0}=1$. The measured coupling parameters are plotted alongside the values calculated with Eqs.~\eqref{eq:alpha} to~\eqref{eq:vfv0}.%

\par Fig.~\ref{fig:scan} confirms that Eqs.~\eqref{eq:alpha} to~\eqref{eq:vfv0} can be used to describe the coupling of the system, for the regimes considered. In Fig.~\ref{fig:scan} a) we observe that the magnetic compression ratio $\alpha$ increases with the driver density and the Mach number, since for these conditions, the magnetic field offers less resistance to the driver, leading to tighter compressions. We can also see that for some of the simulations with low driver densities and low Mach numbers, the driver is not capable of compressing the background, leading to a magnetic decompression with $\alpha<1$.%

\par In Fig.~\ref{fig:scan} b), the measured coupling velocities $v_c$ of the simulations are consistent with Eq.~\eqref{eq:vcv0}. Similarly to the magnetic compression ratio, $v_c$ increases with the driver density and the Mach number. For the simulations with $\alpha<1$, we observe negative coupling velocities, meaning that the driver is pushed back by the background and that it is the background that transfers its energy and momentum to the driver plasma.%

\par Finally, in Fig.~\ref{fig:scan} c), we see that the front velocity $v_f$ increases with the driver density but decreases with the Mach number. For low Mach numbers, we observe small discrepancies between the simulations and Eq.~\eqref{eq:vfv0}. These differences are mostly associated with the difficulty in measuring $v_f$ in this regime, due to low magnetic compressions, \textit{i.e.}, $\alpha\approx1$, and the presence of waves in the background plasma (see Fig.~\ref{fig:basic}). 

\subsection{\label{sub:scan_mass} Dependency on the driver ion mass}

\par
\begin{figure}[!h]
    \centering
    \includegraphics[width=0.95\columnwidth]{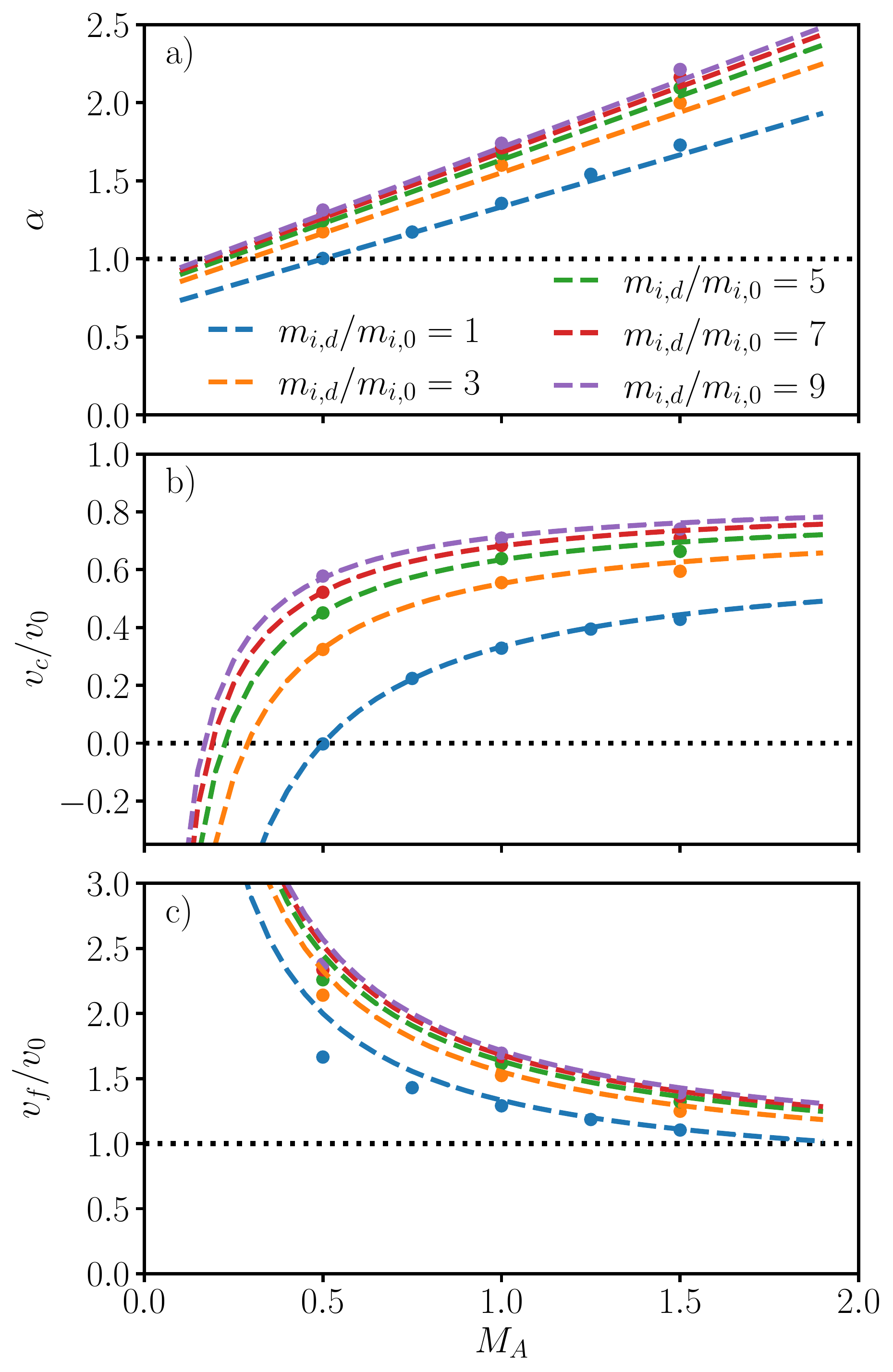}
    \vspace{-2ex}
    \caption{Comparison between the coupling parameters measured in the simulations for a) $\alpha$ with Eq.~\eqref{eq:alpha}, b) $v_c/v_0$ with Eq.~\eqref{eq:vcv0}, and c) $v_f/v_0$ with Eq.~\eqref{eq:vfv0}, for different $M_A$ and $m_{i,d}/m_{i,0}$ values. These simulations considered $n_d/n_0=1$. The coupling parameters measured in the simulations are represented by dots and the analytical expressions by dashed lines. [Associated dataset available at https://zenodo.org/record/7485077 (Ref.~\onlinecite{Zenodo}).]}%
    \label{fig:scan-mass-c} 
\end{figure}

 To continue the validation of Eqs.~\eqref{eq:alpha} to~\eqref{eq:vfv0}, we now compare them in Fig.~\ref{fig:scan-mass-c} to simulations with different driver-to-background ion mass ratios $m_{i,d}/m_{i,0}$ and different Alfvénic Mach numbers $M_A$. The simulations consider $n_d/n_0=1$.%

\par Similarly to Fig.~\ref{fig:scan}, the results of simulations with different ion masses are consistent with Eqs.~\eqref{eq:alpha} to~\eqref{eq:vfv0}. Once again, Eq.~\eqref{eq:vfv0} overestimates the front velocities measured in the simulations for low Mach numbers, mainly due to the difficulty in measuring $v_f$ in these simulations. After confirming that we can use Eqs.~\eqref{eq:alpha} to~\eqref{eq:vfv0} to estimate the coupling parameters for low Mach numbers, we now need to validate the obtained solutions for high Mach numbers.%

\subsection{\label{sub:scan_high} Solutions for high Mach numbers}

\par To study the coupling for higher Mach numbers, we performed additional simulations with $2 \leq M_A \leq 10$. To ensure that a quasi-steady-state was observed in these regimes and that we could measure $\alpha$ with a sufficiently large compressed background region, we considered longer plasmas for these new simulations, namely $L_y=120\ d_i$ for the driver and $L_B=300\ d_i$ for the background.%

\par
\begin{figure}[!h]
    \centering
    \includegraphics[width=0.95\columnwidth]{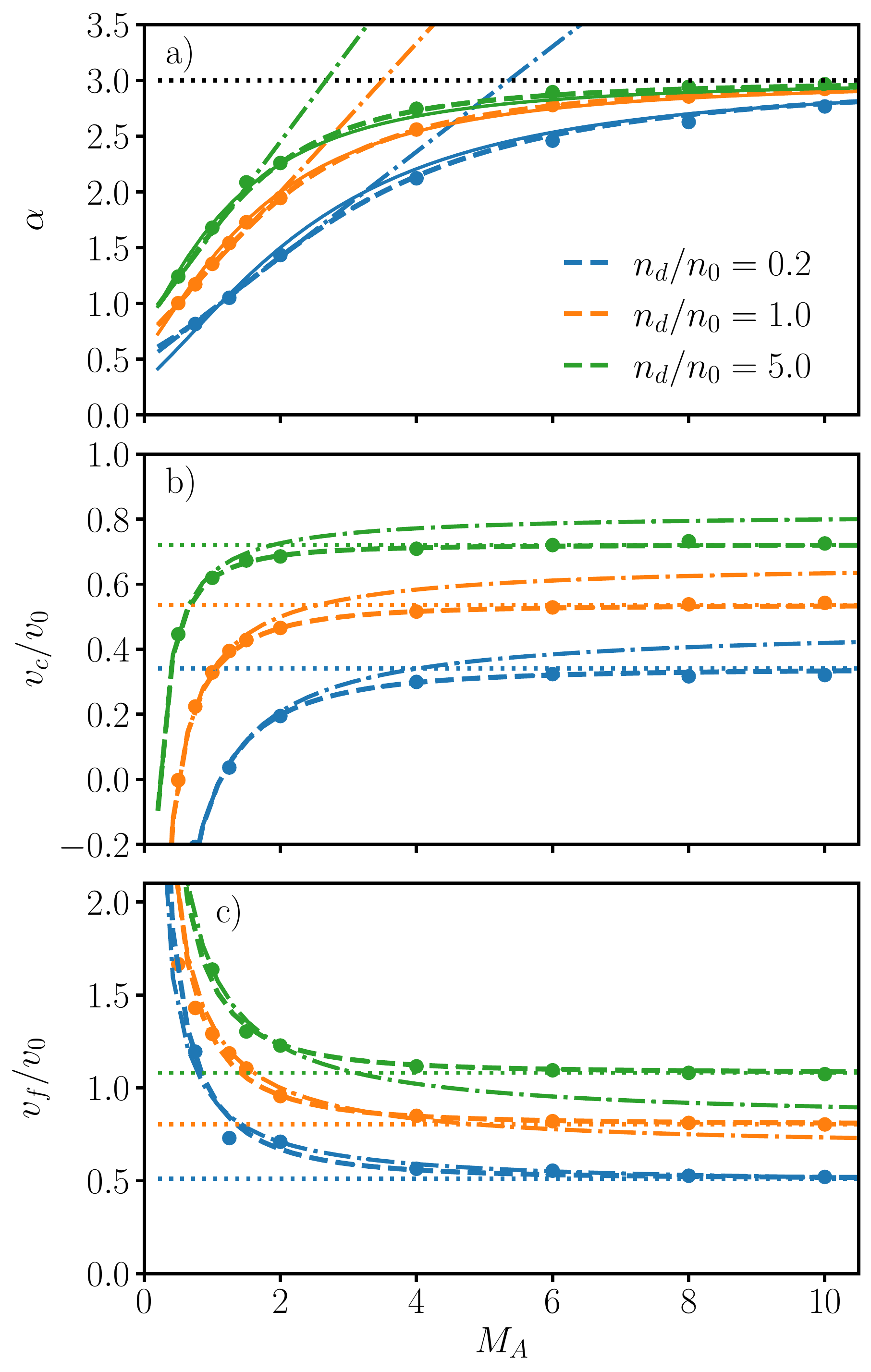}
    \vspace{-2ex}
    \caption{Comparison between the coupling parameters a) $\alpha$, b) $v_c/v_0$, and c) $v_f/v_0$, measured in the simulations and calculated with the solutions obtained in Sec.~\ref{sec:derivation}. The simulations with $M_A<2$ consider $L_y=5\ d_i$ and $L_B=20\ d_i$, while the simulations with $M_A\geq 2$ consider $L_y=120\ d_i$ and $L_B=300\ d_i$. The simulations are represented by dots and considered $m_{i,d}/m_{i,0}=1$. The dashed lines correspond to the solutions obtained by numerically solving Eqs.~\eqref{eq:pressure-balance}, \eqref{eq:alphavcvf}, and~\eqref{eq:large}. The dash-dotted lines correspond to the solutions for low Mach numbers, calculated with Eqs.~\eqref{eq:alpha} to~\eqref{eq:vfv0}, and the dotted lines to the solutions for high Mach numbers, calculated with Eqs.~\eqref{eq:alpha-high} to~\eqref{eq:vfv0-high}. The solid lines in a) correspond to the analytical solutions for $\alpha$ when using Eq.~\eqref{eq:large}, $M_B=M_Av_f/v_0$, and Eq.~\eqref{eq:vfv0}. All solutions considered $\gamma=2$.}
    \label{fig:scan-high} 
\end{figure}

 Fig.~\ref{fig:scan-high} compares the different solutions obtained for the coupling parameters $\alpha$, $v_c/v_0$, and $v_f/v_0$, with the values measured in the simulations, for multiple Mach numbers $M_A$, and the density ratios $n_d/n_0=0.2$, $1$, and $5$. The simulations represented considered $m_{i,d}/m_{i,0}=1$. Since the background magnetic field is aligned along $z$, the heat flow should only be negligible in the $x$ and $y$ directions. As a result, the solutions presented in Fig.~\ref{fig:scan-high} consider $\gamma=2$ for the perpendicularly magnetized background plasma~\cite{Pudovkin1997,Pudovkin1999,Kim2021}.

\par As we can observe in Fig.~\ref{fig:scan-high}, the coupling parameters measured in the simulations are consistent with the solutions obtained by numerically solving Eqs.~\eqref{eq:pressure-balance}, \eqref{eq:alphavcvf}, and~\eqref{eq:large}. As expected, Eqs.~\eqref{eq:alpha} to~\eqref{eq:vfv0} are also consistent with the numerical solutions for low Mach numbers, while Eqs.~\eqref{eq:alpha-high} to~\eqref{eq:vfv0-high} are consistent for high Mach numbers. Unlike the analytical solution for $\alpha$ for low Mach numbers, which corresponds to Eq.~\eqref{eq:alpha}, the numerical solution predicts the saturation of the compression ratio for high Mach numbers. For $\gamma=2$ and $M_A\gg1$, we have $\alpha\approx 3$.%

\par In Fig.~\ref{fig:scan-high} c) we also observe that Eq.~\eqref{eq:vfv0}, the analytical solution for the front velocity $v_f$ and low Mach numbers $M_A$, is similar to the numerical solution and the values measured in the simulations, in particular, for the lower density ratios. If we then calculate $M_B=M_Av_f/v_0$ with Eq.~\eqref{eq:vfv0}, such that
\begin{equation}
    M_B\equiv \frac{v_f}{v_A} \approx \frac{1+M_A}{1+R_n}\ ,
    \label{eq:MB}
\end{equation}
and replace it in Eq.~\eqref{eq:large}, we obtain an additional analytical solution for $\alpha$ that works for all Mach number regimes. As we can see in Fig.~\ref{fig:scan-high} a) with the solid lines, this new analytical solution is a good approximation to the values measured in the simulations and in the numerical solutions.%

\par After confirming that we can describe the coupling between the plasmas with the initial parameters of the system, both analytically and numerically, we can now use the obtained solutions to evaluate other characteristics of the system.%

\subsection{\label{sub:stopping_distance} Stopping distance of the magnetic cavity}

\par Figs.~\ref{fig:scan}, \ref{fig:scan-mass-c}, and~\ref{fig:scan-high} showed that we can use Eqs.~\eqref{eq:alpha} to~\eqref{eq:vfv0} and Eqs.~\eqref{eq:alpha-high} to~\eqref{eq:vfv0-high} to describe how the system evolves over time, for low and high Mach numbers, respectively. These equations are also useful to obtain other parameters, such as the stopping distance $L_{\textrm{stop}}$, \textit{i.e.}, the maximum distance that the magnetic cavity can travel through the background region, during the main interaction of the system.%

\par The reflection time of the magnetic cavity also corresponds to the reflection time of the driver by the background plasma. The magnetic cavity travels a distance $L_\textrm{stop}$ through the background region, with velocity $v_c$, while the driver, with a length $L_{y}$, travels with a velocity $v_0-v_c$, relative to the driver-background boundary. The stopping distance can then be described by%
\begin{equation}
    L_\textrm{stop} = L_{y}\frac{v_c}{v_0-v_c} \ .
    \label{eq:stopping-radius}%
\end{equation}
\noindent Contrary to previously derived expressions for the stopping distance that only consider the driver energy transfer to the magnetic field, for sub-Alfvénic regimes~\cite{Zakharov1999,Ripin1993,Clark2013,Clark2014,Behera2021}, or that only consider the driver energy transfer to the background kinetic energy, for super-Alfvénic regimes~\cite{Winske2007,Clark2013,Clark2014}, Eq.~\eqref{eq:stopping-radius} considers both the background kinetic and magnetic energy transfers and the energy of the reflected driver particles. By using Eq.~\eqref{eq:vcv0} for low Mach numbers, the stopping distance becomes
\begin{equation}
    L_\textrm{stop} = L_{y}\frac{M_A-R_n}{R_n(1+M_A)} \ .
    \label{eq:stopping-radius2}%
\end{equation}
\par
\begin{figure}[!h]
    \centering
    \includegraphics[width=0.95\columnwidth]{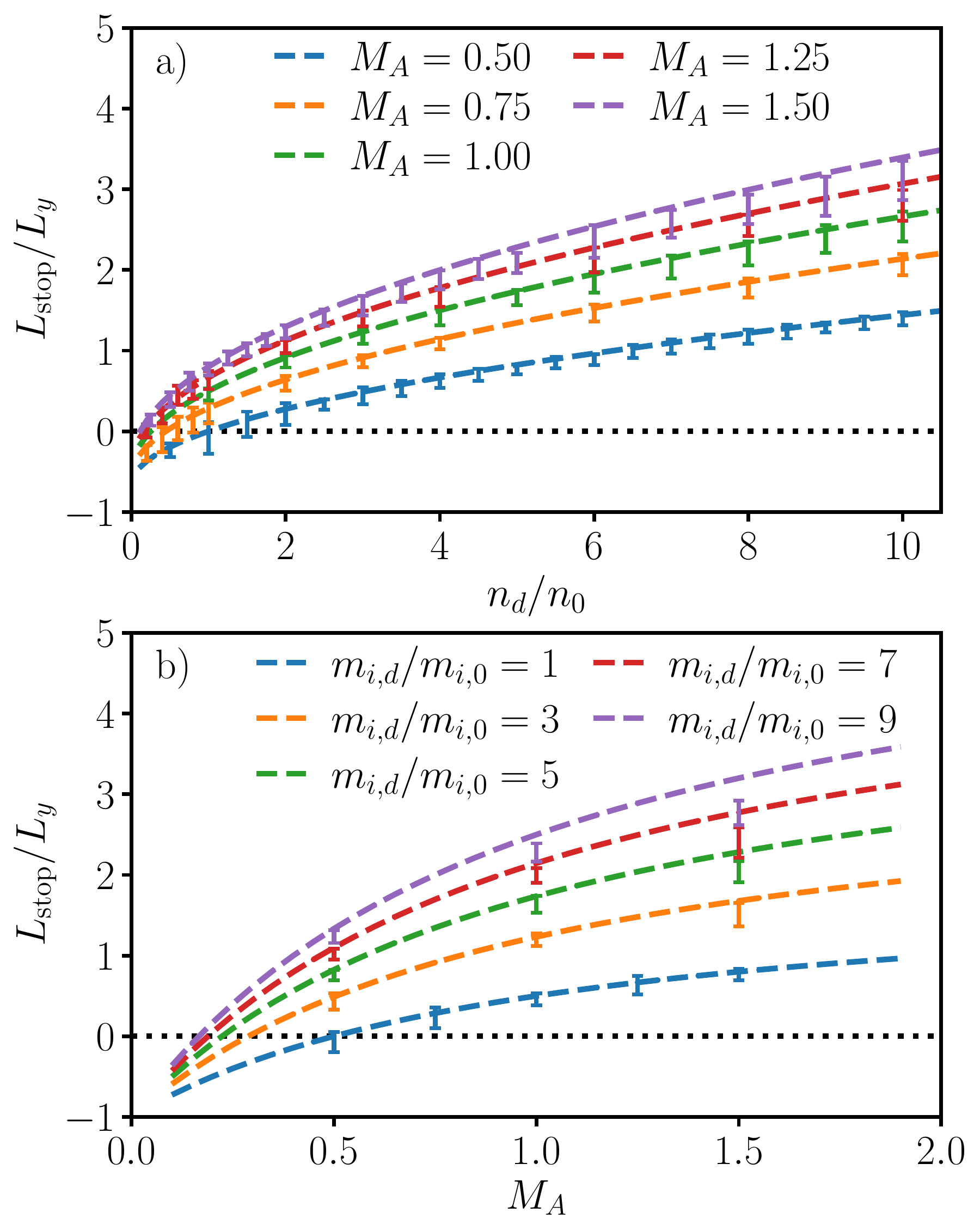}
    \vspace{-2ex}
    \caption{Comparison between the stopping distance $L_{\textrm{stop}}$ measured for multiple simulations, with the uncertainty represented by the errorbars, and calculated with Eq.~\eqref{eq:stopping-radius2}, by the dashed lines. The errorbars account for the non-sharp magnetic cavity reflection in some simulations. a) Scan for different driver-to-background density ratios $n_d/n_0$ and Alfvénic Mach numbers $M_A$, with $m_{i,d}/m_{i,0}=1$. b) Scan for different $M_A$ and driver-to-background ion mass ratios $m_{i,d}/m_{i,0}$, with $n_d/n_{0}=1$. [Associated dataset available at https://zenodo.org/record/7485077 (Ref.~\onlinecite{Zenodo}).]}%
    \label{fig:scan-stopping} 
\end{figure}

 Fig.~\ref{fig:scan-stopping} compares Eq.~\eqref{eq:stopping-radius2} with the measured stopping distance $L_\textrm{stop}$ in the simulations. Fig.~\ref{fig:scan-stopping} a) shows the results for different driver-to-background density ratios $n_d/n_0$ and different Alfvénic Mach numbers $M_A$, with $m_{i,d}/m_{i,0}=1$, and Fig.~\ref{fig:scan-stopping} b) for different driver-to-background ions mass ratios $m_{i,d}/m_{i,0}$ and $M_A$, with $n_d/n_0=1$.%

\par Fig.~\ref{fig:scan-stopping} shows good agreement between Eq.~\eqref{eq:stopping-radius2} and the stopping distances measured in the simulations. For high Mach numbers, densities, and mass ratios, Eq.~\eqref{eq:stopping-radius2} starts to overestimate the stopping distances. As we observed in Fig.~\ref{fig:scan} b) and Fig.~\ref{fig:scan-mass-c} b), for these parameters, the coupling velocity is close to $v_0$. Since the stopping distance is proportional to $v_c/(v_0-v_c)$, small discrepancies between Eq.~\eqref{eq:vcv0} and the simulations may lead to large differences in the $L_\textrm{stop}$ values, for these conditions.%

\par
\begin{figure}[!h]
    \centering
    \includegraphics[width=0.95\columnwidth]{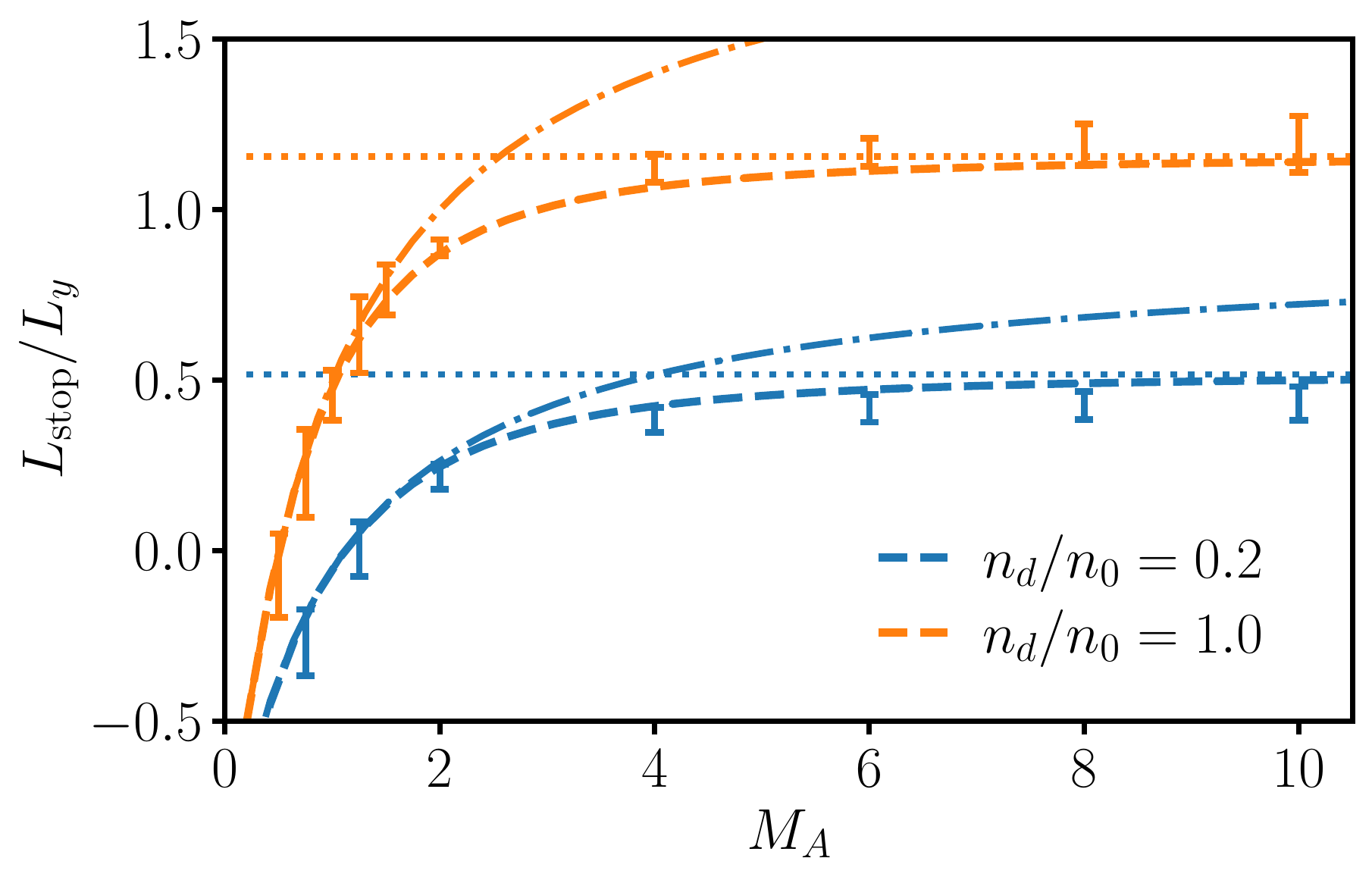}
    \vspace{-2ex}
    \caption{Comparison between the stopping distance $L_{\textrm{stop}}$ measured for multiple simulations, with the uncertainty represented by the errorbars, and Eq.~\eqref{eq:stopping-radius}. The errorbars account for the non-sharp magnetic cavity reflection and the formation of a ``foot'' structure in the driver-background interface in some simulations. The dashed lines correspond to values calculated with Eq.~\eqref{eq:stopping-radius} and the numerical solutions of $v_c$, the dash-dotted lines to the low Mach number solution of Eq.~\eqref{eq:stopping-radius2}, and the dotted lines to the high Mach number solution of Eq.~\eqref{eq:stopping-radius2-high}. The simulations considered different driver-to-background density ratios $n_d/n_0$ and Alfvénic Mach numbers $M_A$, with $m_{i,d}/m_{i,0}=1$. The simulations used a driver length of $L_y=5\ d_i$, for $M_A<2$, and $L_y=120\ d_i$, for $M_A\geq 2$.}%
    \label{fig:scan-stopping-high} 
\end{figure}

 For high Mach numbers, and using Eq.~\eqref{eq:vcv0-high}, the stopping distance is given by%
\begin{equation}
    L_\textrm{stop} = \frac{L_{y}}{R_n\sqrt{1+\gamma}} \ .
    \label{eq:stopping-radius2-high}%
\end{equation}
To also validate the solutions for high Mach numbers, Fig.~\ref{fig:scan-stopping-high} compares the values measured in the simulations with the obtained solutions for the different Mach number regimes. The simulations consider $m_{i,d}/m_{i,0}=1$ and the density ratios $n_d/n_0=0.2$ and $1$. 

\par Similarly to Fig.~\ref{fig:scan-high}, Fig.~\ref{fig:scan-stopping-high} shows that the stopping distance $L_\textrm{stop}$ calculated with Eq.~\eqref{eq:stopping-radius} and the numerical solutions of $v_c$ are consistent with the values measured in the simulations, from low to high Mach numbers. Additionally, Eq.~\eqref{eq:stopping-radius2} is consistent with the simulations for low Mach numbers, while Eq.~\eqref{eq:stopping-radius2-high} is consistent with the simulations for high Mach numbers. In some of the high $M_A$ simulations, some ``foot'' structures were observed in the driver-background interface, leading to some uncertainty in the measured values of the stopping distance.%

\section{\label{sec:applications} Applications to experiments and simulations}

\par Eqs.~\eqref{eq:alpha} to~\eqref{eq:vfv0} relate the different coupling parameters with the initial conditions of the system, for the low Mach numbers expected for the experiments with ion-scale magnetospheres. To validate the developed model, we now compare it with results from experiments with laser-driven plasmas and magnetized background plasmas. In particular, we compare the model with runs from LAPD experiments that observed strong coupling between the two plasmas.%

\par The data chosen for these experiments are shown in Table~\ref{tab:table}. The typical laser-driven plasmas produced in the laboratory are not uniform in density and velocity as our coupling model considers. However, under small scales of the driver expansion and for the strong coupling regimes of the selected runs, some features remain similar to our 1D simulations, such as constant coupling and front velocities, and a plateau region on the magnetic field compression. In these conditions, we can partially apply the coupling study to experiments.%
\begin{table}
\caption{\label{tab:table} Coupling parameters of LAPD experiments with laser-driven and magnetized background plasmas~\cite{Schaeffer2015,Everson2016}. We selected runs with $M_A\sim1$ and strong coupling regimes, \textit{i.e.}, runs with magnetic cavities with approximately no magnetic field inside, a plateau region with a strong magnetic field compression, and constant coupling and front velocities for sufficiently long times. We also selected the run shown in Fig.~\ref{fig:experiment} a) for comparison~\cite{Schaeffer2022}.}
\begin{ruledtabular}
\begin{tabular}{ccccccc}
Run&$v_A$ (km/s)&$v_c$ (km/s)&$v_f$ (km/s)& $\alpha$ &$v_c/v_A$ & $v_f/v_A$\\
\hline
Run4 2013~\cite{Everson2016} & 280 & 164 & 250 & 1.6 & 0.6 & 0.9\rule{0pt}{2.6ex} \\
Run3 2015~\cite{Schaeffer2015} & 189 & 260 & 440 & 2.0 & 1.4 & 2.3 \\
Mini Mag.~\cite{Schaeffer2022} & 378 & 135 & 380 & 1.3 & 0.4 & 1.0 \\
\end{tabular}
\end{ruledtabular}
\end{table}
\par Since the experimental driver is not uniform and we do not have accurate measurements of the density, length, and velocity of the plasma, we cannot properly calculate the corresponding $R_n$ quantity in these experiments. With Eqs.~\eqref{eq:alpha} to~\eqref{eq:vfv0}, we obtain $v_f/v_A=\alpha$, and $v_c/v_A=\alpha-1$. We can use these relations to verify experimentally the validity of our model.%

\par For the first and second runs in Table~\ref{tab:table}, we have an average compression of $\alpha>1.5$, leading to strong coupling between the plasmas. For the first run, we observe $v_c/v_A\approx \alpha-1$, as expected from the coupling model. However, we also observe $v_f/v_A<\alpha$. For the second run, we observe $v_f-v_c\approx v_A$, but also $v_f/v_A>\alpha$. These differences from the coupling model may have emerged from the typical deceleration of the cavity and compression expansions observed in experiments~\cite{Ripin1993,Schaeffer2015,Everson2016,Winske2019}, and from the difficulty in measuring precisely some of the coupling parameters from the available data.%

\par In the last row of Table~\ref{tab:table}, we have the coupling parameters for the experiments with ion-scale magnetospheres~\cite{Schaeffer2022}, for the no dipole case, represented in Fig.~\ref{fig:experiment} a). Since $\alpha<1.5$, this run has weaker coupling between the plasmas than the two previous cases~\cite{Schaeffer2015,Everson2016}. We observe $v_c/v_A\approx \alpha-1$, but $v_f/v_A<\alpha$. In this run, the plasmas are short and do not interact with each other for enough time to observe a plateau region in the magnetic compression.%

\par We can also apply the coupling study to estimate the spatial and temporal scales of laboratory ion-scale magnetospheres in experimental~\cite{Schaeffer2022} and numerical~\cite{Cruz2022} studies. As discussed in Sec.~\ref{sec:review}, these systems consider a driver plasma and magnetized background plasma, with a dipolar magnetic field centered in the background. The pressure balance in Eq.~\eqref{eq:pressure-mag} that describes the magnetopause current observed in Fig.~\ref{fig:simulation} b) depends on the parameters of the system after the driver and background plasmas interact with each other and start moving towards the dipole. 
Since these parameters depend on $\alpha$, $v_c$, and $v_f$, which can be calculated with Eqs.~\eqref{eq:alpha} to~\eqref{eq:vfv0} for the expected low Mach numbers in the laboratory~\cite{Schaeffer2022}, we can use the coupling study to obtain more accurately the pressure balances described by Eq.~\eqref{eq:pressure-mag} that determine the standoff locations in ion-scale magnetospheres. 

\par Furthermore, to observe the magnetopause under this setup, we must make sure that the driver has sufficient energy to push enough background plasma toward the dipole. With Eq.~\eqref{eq:stopping-radius}, we can determine how further can the driver plasma travel, and with Eq.~\eqref{eq:pressure-mag}, we can determine the effective size of the magnetosphere. Using these two quantities, we can estimate if the driver has enough energy to ensure the observation of a laboratory ion-scale magnetosphere, in the experiments and the simulations.


\section{\label{sec:conclusions} Discussion and conclusions}

\par In recent experiments on the Large Plasma Device (LAPD) at UCLA, ion-scale magnetospheres were performed in the laboratory by driving a laser-produced plasma into a dipolar magnetic field embedded in a uniformly magnetized plasma. Under this configuration, the laser-driven and background plasmas first interact with each other before interacting with the strong magnetic field of the dipole. For the experimental and numerical analysis of laboratory ion-scale magnetospheres, and for the design of future experiments that involve fast plasmas moving toward magnetized plasmas, it is necessary to understand this interaction.%

\par In this paper, we derived analytical expressions for magnetic field parameters that describe the coupling between an unmagnetized driver plasma and a perpendicularly magnetized background plasma. These expressions were then compared with 1D particle-in-cell (PIC) simulations for multiple densities, ion masses, and magnetic field values. For the cold plasmas, and uniform density and velocity profiles considered, the expressions were consistent with simulations. These expressions allow us to (i) evaluate the coupling between the plasmas, (ii) estimate initial quantities from simple magnetic field diagnostics, and (iii) calculate the spatial and temporal scales of these systems.%


\par For the ideal plasmas considered, the simulations reached a near steady-state condition, where the coupling parameters --- the average magnetic compression ratio and the velocities of the magnetic cavity and of the magnetic compression --- remain constant. These quantities describe the coupling between the plasmas and increase with higher driver-to-background density and ion mass ratios. The compression ratio and the cavity velocity also increase with the Alfvénic Mach number, while the compression velocity decreases. Additionally, for some parameters, the driver plasma does not have enough momentum to push the background forward.%

\par From conservation arguments, we obtained analytical expressions and numerical solutions for the coupling parameters, which were consistent with 1D PIC simulations. Since these parameters can be measured from magnetic field diagnostics, they can be used as a benchmark for the initial conditions of these systems. With these expressions, we can also determine other quantities, such as the stopping distance of the magnetic cavity and the magnetopause position  associated with the laboratory ion-scale magnetospheres.%

\par We assumed uniform profiles and long plasmas for the coupling model, and always observed strong coupling and a quasi-steady-state regime in the simulations. In the experiments with laser-driven plasmas and magnetized background plasmas, however, the driver is short, non-uniform, and expanding, and therefore, we do not always observe the same conditions. A complete study of the experimental coupling between the plasmas must consider these characteristics.%

\par In conclusion, we derived analytical expressions for multiple parameters and arbitrary Alfv\'{e}nic flows that describe the coupling between a driven plasma and a magnetized background plasma. These expressions are consistent with results from PIC simulations and can assist in the design of future experiments with driven plasmas and magnetized obstacles. For future works, we intend to explore other regimes and configurations, such as higher ion and electron temperatures, shorter drivers, and non-uniform densities and velocities.%

\begin{acknowledgments}
We acknowledge the support of the European Research Council (InPairs ERC-2015-AdG 695088), FCT (PD/BD/114307/2016, APPLAuSE PD/00505/2012, and UID/FIS/50010/2023), the NSF/DOE Partnership in Basic Plasma Science and Engineering (Award Number PHY-2010248), and PRACE for awarding access to MareNostrum (Barcelona Supercomputing Center, Spain). The simulations presented in this work were performed at the IST cluster (Lisbon, Portugal) and at MareNostrum.
\end{acknowledgments}

\section*{Data Availability Statement}


\par The data that support the findings of this study are openly available in Zenodo at http://doi.org/10.5281/zenodo.7485077, reference number~\onlinecite{Zenodo}.

\appendix

\section{\label{sec:appendix} Electric field of the system}

\par For the collisionless, magnetic pressure dominated ($\beta \equiv 8\pi n_eT_{e}/B^2\ll1$, where $n_e$ and $T_e$ are the electron density and temperature, respectively), and low Mach numbers ($M_A\sim1$) considered, neither collisions nor instabilities effectively transfer momentum and energy between the driver and the perpendicularly magnetized background plasma. For these conditions, the laminar electric field provides the dominant coupling mechanism between the two plasmas~\cite{Everson2016}. Using a hybrid model~\cite{Bondarenko2017}, where the ion species are considered kinetically, and the electron species as a charge-neutralizing fluid, and considering that the magnetic field is mostly defined in the $z$ direction, then the laminar collisionless electric field of the system, for the regimes considered, is approximately given by%
\begin{equation}
    \mathbf{E} \approx -\frac{1}{4\pi en_e}B_z\mathbf{\nabla_\bot} B_z - \frac{1}{en_ec}(\mathbf{J_d}+\mathbf{J_0})\times\mathbf{B_z} -\frac{\nabla p}{en_e}\ .
    \label{eq:bondarenko}
\end{equation}
\noindent In Eq.~\eqref{eq:bondarenko}, $\mathbf{J_j}=Z_jn_j\mathbf{v_j}$ is the current density of the driver ($j=d$) or of the background ($j=0$) plasmas. $Z_j$, $n_j$, and $\mathbf{v_j}$ are the ions' charge, density, and velocity, respectively, for the plasma $j$. With quasi-neutrality, we have $n_e\approx Z_dn_d+Z_0n_0$. The first term in Eq.~\eqref{eq:bondarenko}, $\mathbf{E_1}=-B_z\mathbf{\nabla_\bot}B_z/4\pi en_e$, is primarily defined along the $y$ direction since $|\partial B/\partial x|\ll|\partial B/\partial y|$. The second term, $\mathbf{E_2}=-(\mathbf{J_b}+\mathbf{J_d})\times\mathbf{B_z}/en_ec$, however, is mostly defined in $x$, with $\mathbf{E_2}\approx-(v_y/c)B_z\ \mathbf{x}$, since the ion motions are mostly defined along $y$. The third term $\mathbf{E_3}=-\nabla p/en_e$ is associated with the electron pressure $p$.%

\par
\begin{figure}[!h]
    \centering
    \includegraphics[width=0.95\columnwidth]{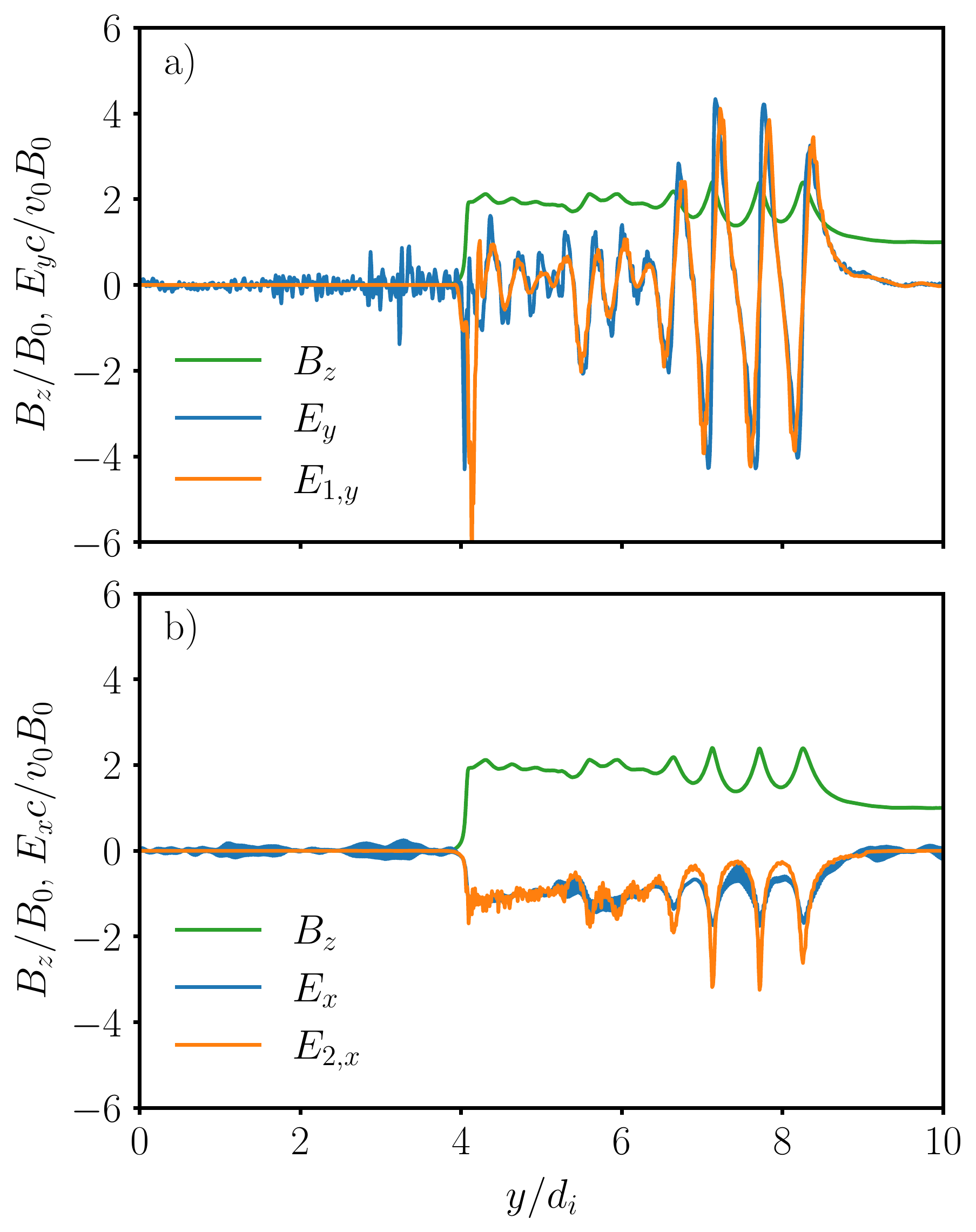}
    \vspace{-2ex}
    \caption{Comparison between the terms of Eq.~\eqref{eq:bondarenko} and the electric field of the simulation with $n_d/n_0 = 2$, $m_{i,d}/m_{i,0}=1$, and $M_A=1.5$, for $t\omega_{ci}\approx 5.0$. The $y$ components of the electric field are shown in a) and the $x$ components in b). $E_y$ and $E_x$ refer to the electric field profiles of the simulation (blue), and $E_1$ and $E_2$ to the first and second terms of Eq.~\eqref{eq:bondarenko} (orange). Both frames also show the magnetic field $B_z$ (green).}%
    \label{fig:laminar-electric-field} 
\end{figure}

 To verify if Eq.~\eqref{eq:bondarenko} correctly describes the electric field of the system, the terms in Eq.~\eqref{eq:bondarenko} and the electric field of the simulation with $n_d/n_0 = 2$, $m_{i,d}/m_{i,0}=1$, and $M_A=1.5$ (previously presented in Fig.~\ref{fig:basic}), are compared in Fig.~\ref{fig:laminar-electric-field}, for $t\omega_{ci}\approx 5.0$.%

\par Fig.~\ref{fig:laminar-electric-field} shows that we can use Eq.~\eqref{eq:bondarenko} to calculate the electric field of these systems. In Fig.~\ref{fig:laminar-electric-field} a), we see that the $y$ component of the electric field can be expressed by $\mathbf{E_1}$, which depends on the magnetic field gradient. Within the interface that separates the magnetic cavity from the compressed magnetic field, we observe a significant negative electric field. This electric field is responsible for the reflection of the driver ions back to the upstream region.%

\par Fig.~\ref{fig:laminar-electric-field} b) also shows that the $x$ component of the electric field can be described by $\mathbf{E_2}$. In the background region, $E_x$ is always negative and approximately $-(v_y/c)B_z$, in agreement with Ohm's law, $\mathbf{E}+\mathbf{v}\times\mathbf{B}/c = 0$. For the low Mach numbers and low $\beta$ considered, $\mathbf{E_3}$ is typically negligible when compared to the electric field terms $\mathbf{E_1}$ and $\mathbf{E_2}$. 

\par We can use Eq.~\eqref{eq:bondarenko} to describe the motion of the particles, in particular, the reflection of the driver particles by the background plasma region. Near this region, unmagnetized driver ions move with velocity $v_0$ against the compressed background magnetic field of average value $\alpha B_0$. The discontinuity region (labeled as $A$ in Fig.~\ref{fig:model}) moves with velocity $v_c$, and the driver ions end up reflected upstream with velocity $v_1$. From Eq.~\eqref{eq:bondarenko}, this field is approximately given by%
\begin{equation}
    E_y \approx - \frac{1}{4\pi en_e}B_z\frac{\partial B_z}{\partial y} - \frac{1}{en_e}\frac{\partial p}{\partial y}\ .
    \label{eq:electric-field}%
\end{equation}
Since $|E_y|\gg |v_xB_z/c|$ for this region, the equation of motion for a reflecting driver ion is%
\begin{equation}
    m_{i,d}\frac{dv'_{i,d,y}}{dt'} = m_{i,d}\frac{dv'_{i,d,y}}{dy'}v'_{i,d,y} = Z_deE_y \ ,
    \label{eq:inter1}
\end{equation}
\noindent where $v'_{i,d,y}\equiv v_{i,d,y}-v_c$ is the driver ion velocity in the reference frame of the discontinuity. In this frame, the driver ions have initial velocity $v'_{i,d,y}=v_0-v_c$ and start to be reflected upstream when $v'_{i,d,y}=0$. By integrating Eq.~\eqref{eq:inter1}, and considering $p_2$ as the average electron pressure of the compressed background plasma, we obtain%
\begin{align}
    \int^0_{v_0-v_c} v_{i,d,y}'\ dv_{i,d,y}' = &- \frac{Z_d}{4\pi m_{i,d}}\int^{\alpha B_0}_0 \frac{B_z'}{n_e'}dB_z' \nonumber \\ 
    &-\frac{Z_d}{m_{i,d}}\int^{p_2}_0 \frac{1}{n_e'}dp'\ .
    \label{eq:inter2}
\end{align}
We consistently observed a peak in the electron density of $n_e\approx 4\ Z_dn_d$ in the interface driver-background of the simulations. Assuming this value in Eq.~\eqref{eq:inter2}, we end up with
\begin{align}
    \frac{(v_0-v_c)^2}{2} = \frac{1}{4\pi m_{i,d}}\frac{(\alpha B_0)^2}{8n_d} + \frac{1}{m_{i,d}}\frac{p_2}{4n_d}\nonumber \\  \iff 2n_{d}m_{i,d}(v_0-v_c)^2 = p_2+\frac{(\alpha B_0)^2}{8\pi}\ ,
    \label{eq:pressure-avc}
\end{align}
\noindent which corresponds to the pressure balance of Eq.~\eqref{eq:pressure-balance} for $\beta_d\ll 1$, as expected.

\section{\label{app:energy} Energy expressions}

\par To validate the assumption $\alpha\approx M_B\equiv v_f/v_A$ for low Mach numbers considered in Sec.~\ref{sub:jumpB}, we now compare the different energy fluxes terms represented in Eq.~\eqref{eq:energy_conservation-rh} with the total energies measured in multiple simulations. Considering $\phi_{\textrm{d}}$, $\phi_{0}$, $\phi_{\textrm{mag}}$, and $\phi_{\textrm{ele}}$ the energy fluxes of the system associated with the driver plasma, background plasma, magnetic field, and electric field, respectively, in the lab frame, over a time $\delta t$ and transverse area $a_t$, we must have, due to energy conservation%
\begin{align}
  \Phi_{\textrm{d}}+\Phi_{0}+\Phi_{\textrm{mag}}+\Phi_{\textrm{ele}} = 0 \ .
  \label{eq:energy_conservation}
\end{align}
Since $|\Phi_{\textrm{ele}}/\Phi_{\textrm{mag}}| \sim (v_0/c)^2 \ll 1$ (see Appendix~\ref{sec:appendix}), $\Phi_{\textrm{ele}}$ can be neglected in Eq.~\eqref{eq:energy_conservation}.%

\par In Fig.~\ref{fig:model}, we observe that the driver consists of two populations, with velocity $v_0$ and $v_1$. The energy flux of each population can be calculated by multiplying the kinetic energy of each ion with the rate of the number of ions. Recalling from Sec.~\ref{sub:jumpA} that $n_d'=n_d$ and $v_1=2v_c-v_0$, the interface driver-background travels with velocity $v_c$, and assuming cold plasmas ($\beta_1,\beta_d\ll1$), we obtain that the driver energy flux in the lab frame can be calculated with%
\begin{align}
  \Phi_{\textrm{d}} &= -\frac{1}{2}m_{i,d}v_0^2n_d(v_0-v_c) + \frac{1}{2}m_{i,d}v_1^2n_d'(v_c-v_1) \nonumber \\ &= -2n_dm_{i,d}(v_0-v_c)^2v_c\ .
  \label{eq:energy_driver}%
\end{align}
Unlike the driver, the background plasma is located in two different regions in Fig.~\ref{fig:model}. In the magnetic compression region, the average kinetic energy flow of each background ion is $m_{i,0}v_c^2/2$, and the density is $n_0'=\alpha n_0$. In the unperturbed background region, the background plasma has no flow velocity. For low Mach numbers and cold plasmas, we can ignore the contribution of the compressed background electron pressure $p_2$. Since the back and front boundaries of the compression region travel with velocities $v_c$ and $v_f$, respectively, we can then express the background energy flux associated with the plasmas flow as%
\begin{equation}
  \Phi_{0} = \frac{1}{2}m_{i,0}v_c^2n_{0}'(v_f-v_c) = \frac{1}{2}m_{i,0}v_c^3n_{0}\frac{\alpha}{\alpha-1} \ .
  \label{eq:energy_bg1}%
\end{equation}
\par Finally, we need an expression for the magnetic energy flux. The magnetic compression region has an average magnetic field of $\alpha B_0$ and increases its length at a velocity $v_f-v_c$, while the unperturbed background region has a magnetic field $B_0$ and a length that decreases at a velocity $-v_f$. The magnetic energy flux is thus%
\begin{align}
  \Phi_{\textrm{mag}} &= \frac{(\alpha B_0)^2}{8\pi}(v_f-v_c) - \frac{B_0^2}{8\pi}v_f = \frac{B_0^2}{8\pi}\alpha v_c \ .  
  \label{eq:energy_magnetic}%
\end{align}
\par To validate the previous expressions for the energy fluxes, we performed multiple simulations with different Alfvénic Mach numbers $M_A$ and driver-to-background density ratios $n_d/n_0$, with $m_{i,d}/m_{i,0}=1$. For each simulation, we measured the coupling parameters $\alpha$ and $v_c$ from the magnetic field data, and then, with these values, we calculated the theoretical energy fluxes with Eqs.~\eqref{eq:energy_driver} to~\eqref{eq:energy_magnetic}.%

\par
\begin{figure}[!h]
    \centering
    \includegraphics[width=0.95\columnwidth]{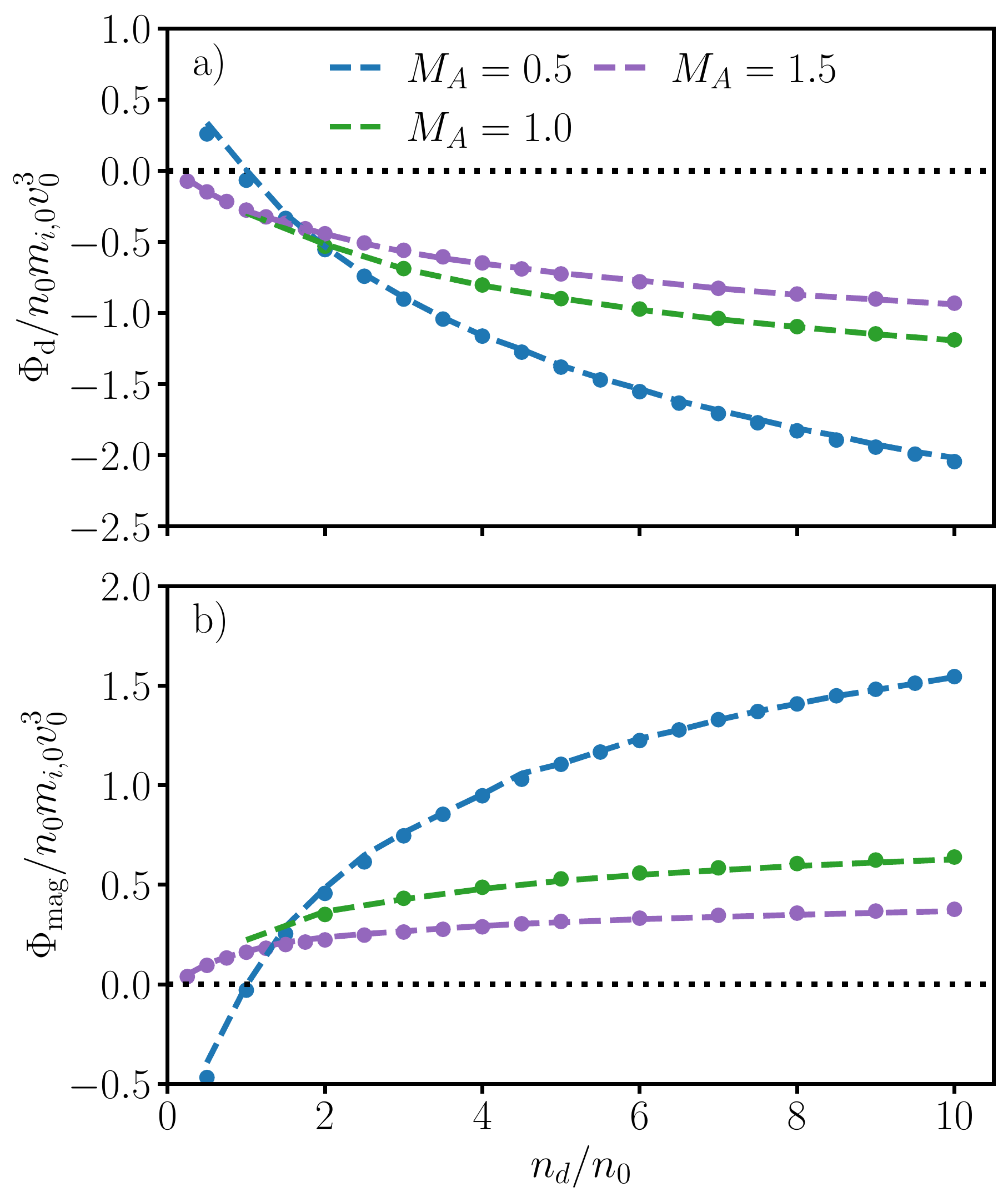}
    \vspace{-2ex}
    \caption{Energy fluxes of a) the driver ions $\Phi_\textrm{d}$ and b) the magnetic field $\Phi_\textrm{mag}$, for multiple simulations with different Alfvénic Mach numbers $M_A$ and density ratios $n_d/n_0$. We considered $m_{i,d}/m_{i,0}=1$. The energy fluxes calculated from the total energy are represented by scatter points, while the energy fluxes calculated with Eqs.~\eqref{eq:energy_driver} and~\eqref{eq:energy_magnetic}, and with the measured $\alpha$ and $v_c$ values, are connected by dashed lines. [Associated dataset available at https://zenodo.org/record/7485077 (Ref.~\onlinecite{Zenodo}).]}%
    \label{fig:energy-verification}%
\end{figure}

 In Fig.~\ref{fig:energy-verification}, the calculated energy fluxes of a) the driver plasma $\Phi_\textrm{d}$ and b) the magnetic field $\Phi_\textrm{mag}$ are compared to the correspondent energy fluxes obtained from the variation of the total energies in the simulations. Each quantity was measured in the quasi-steady-state regime of the simulations.%

\par Fig.~\ref{fig:energy-verification} confirms that Eqs.~\eqref{eq:energy_driver} and~\eqref{eq:energy_magnetic} can be used to describe the driver and magnetic energy fluxes of the system, for the regimes considered. In Fig.~\ref{fig:energy-verification} a) we observe that the flux of energy lost by the driver increases for higher driver densities and lower Mach numbers, leading to a more efficient energy transfer from the driver to the background. Similarly, in Fig.~\ref{fig:energy-verification} b), the magnetic energy flux is larger for denser drivers and lower Mach numbers.%

\par In Fig.~\ref{fig:energy-verification}, we also observe that in the simulation with $M_A = 0.5$ and $n_d / n_0 = 0.5$, it is the magnetic field that transfers energy to the driver plasma. This simulation corresponds approximately to the case where the initial magnetic pressure $B_0^2/8\pi$ is larger than the driver's ram pressure $2n_dm_{i,d}v_0^2$ (\textit{i.e.}, $R_n>M_A$), and so, the driver is pushed back by the background, leading to a negative coupling velocity.%

\par
\begin{figure}[!h]
    \centering
    \includegraphics[width=0.95\columnwidth]{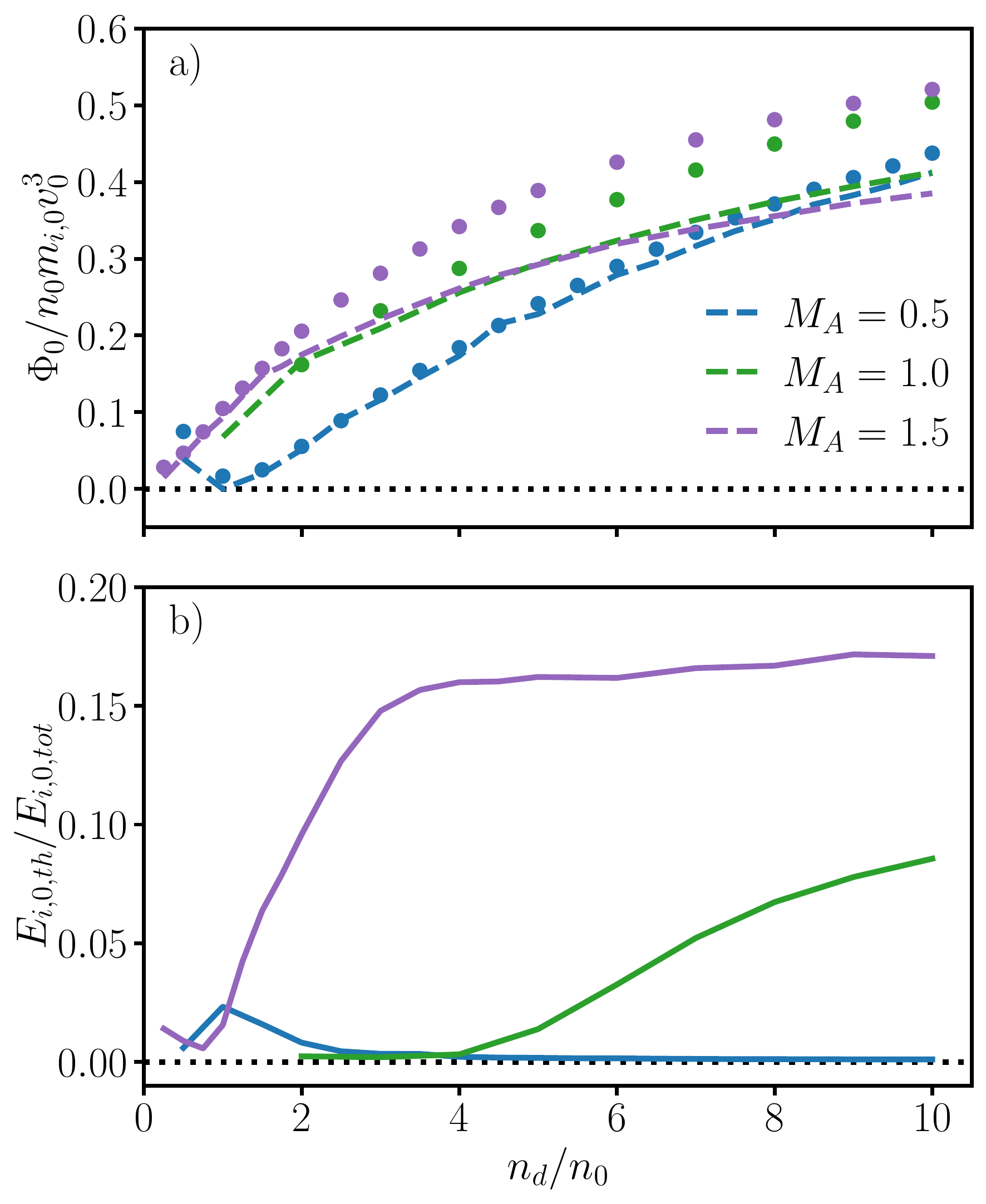}
    \vspace{-2ex}
    \caption{a) Energy flux of the background plasma $\Phi_0$, for multiple simulations with different Alfvénic Mach numbers $M_A$ and density ratios $n_d/n_0$. We considered $m_{i,d}/m_{i,0}=1$. The fluxes calculated from the total energy are represented by scatter points, while the energy fluxes calculated from Eq.~\eqref{eq:energy_bg1}, and with the measured $\alpha$ and $v_c$ values, are connected by dashed lines. b) Fraction of thermal energy to total energy of the background ions, in the final stage of the main interaction, for each simulation. [Associated dataset available at https://zenodo.org/record/7485077 (Ref.~\onlinecite{Zenodo}).]}%
    \label{fig:energy-verification2} 
\end{figure}

 Fig.~\ref{fig:energy-verification2} a) compares the background energy flux $\Phi_{0}$ calculated with Eq.~\eqref{eq:energy_bg1} with the energy flux obtained from the total energy of the simulations. As we can see, Eq.~\eqref{eq:energy_bg1} is more consistent with the simulations for low Mach numbers, than for high Mach numbers and driver densities.%

\par For high Mach numbers, instabilities start to form in the background plasma. This leads to an increase in the thermal energy of the plasma, which starts to have an important role in the energy partition of the system. Since Eq.~\eqref{eq:energy_bg1} neglects instabilities and thermal effects, it underestimates the energy flux of the background plasma, as we observe in Fig.~\ref{fig:energy-verification2} a). Fig.~\ref{fig:energy-verification2} b) shows the average ratio of thermal energy of the background ions $E_{i,0,th}$ to their total energy $E_{i,0,tot}$, near the final stage of the main interaction in the simulations. As expected, we observe that, for high Mach numbers and driver densities, a significant percentage of the background ions' energy is thermal energy, while for low Mach numbers, the thermal energy is negligible.

\par The thermal effects of the ions in the compressed background can be estimated with the pressure $p_2$. If we assume Eqs.~\eqref{eq:alpha} and~\eqref{eq:vfv0} for $\alpha$ and $v_f$, respectively, and apply Eq.~\eqref{eq:beta2}, we obtain that the ratio of the plasma pressure to magnetic pressure in the compressed background region is given by
\begin{equation}
    \frac{8\pi p_2}{(\alpha B_0)^2}=\frac{\beta_2}{\alpha^2}\approx\left(\frac{\alpha-1}{\alpha}\right)^2=\left(\frac{v_c}{v_f}\right)^2 \ .
    \label{eq:ratio_p2}
\end{equation}
As we can see in Fig.~\ref{fig:scan}, the coupling velocity $v_c$ increases with the Mach number $M_A$ and the density ratio $n_d/n_0$, while the front velocity $v_f$ decreases with $M_A$ but increases with $n_d/n_0$. As a result, the ratio in Eq.~\eqref{eq:ratio_p2} decreases for low Mach numbers. In this regime, we can then ignore the $p_2$ term in the energy conservation. 

\par By applying Eqs.~\eqref{eq:energy_driver} to~\eqref{eq:energy_magnetic} in Eq.~\eqref{eq:energy_conservation}, it is possible to show that $\alpha \approx v_f/v_A=M_B$ for low Mach numbers. Additionally, we obtain from Eq.~\eqref{eq:large} for $\gamma=2$ and $\beta_1\ll1$ that $\alpha=M_B$ for $M_B=0$, $1$, and $\gamma/(\gamma-1)=2$. As a result, we observe that Eq.~\eqref{eq:large} follows $\alpha\approx M_B$ for low Mach numbers $M_B<2$.%

\par From Eqs.~\eqref{eq:energy_driver} and~\eqref{eq:energy_bg1}, we also obtain that the ratio of energy transferred from the driver plasma to the background plasma is
\begin{equation}
    -\frac{\Phi_{0}}{\Phi_{\textrm{d}}} = \frac{M_A-R_n}{M_A+1} = \frac{v_c}{v_f} \ .
    \label{eq:energy_transfer}
\end{equation} 
The ratio between the velocities of the magnetic cavity and of the magnetic compression is then a direct tool to evaluate the efficiency of the energy transfer from the driver to the background plasmas.%

\nocite{*}

\bibliography{aipsamp}

\end{document}